\documentclass[11pt]{article}	
\pdfoutput=1
\usepackage{jcappub}

\usepackage{natbib}

\usepackage{amsmath,amsfonts,amssymb}
\usepackage{setspace}
\usepackage[T1]{fontenc} % if needed
\usepackage{subfigure}
\usepackage{wrapfig}
\usepackage{graphicx}
\usepackage{longtable}

\usepackage[normalem]{ulem}
\usepackage{color}

\title{The Primordial Inflation Explorer (PIXIE): 
Mission design and science goals}

\author[a,1]{Alan Kogut%
\note{Corresponding author.}}
\author[b]{Nabila Aghanim}
\author[c]{Jens Chluba}
\author[d]{David T. Chuss}
\author[e]{Jacques Delabrouille}
\author[f]{Cora Dvorkin}
\author[a,g]{Dale Fixsen}
\author[h,e]{Shamik Ghosh}
\author[i]{Brandon S. Hensley}
\author[j,k]{J. Colin Hill}
\author[l]{Bruno Maffei}
\author[m,k]{Anthony R. Pullen}
\author[c]{Aditya Rotti}
\author[n]{Alina Sabyr}
\author[a]{Eric R. Switzer}
\author[o,p]{Leander Thiele}
\author[a]{Edward J. Wollack}
\author[q]{Ioana Zelko}

\affiliation[a]{NASA Goddard Space Flight Center, 8800 Greenbelt Road, Greenbelt, MD 20771, USA}
\affiliation[b]{Universit\'e Paris-Saclay, CNRS, Institut d'Astrophysique Spatiale, B\^atiment 121, 91405 Orsay, France}
\affiliation[c]{Jodrell Bank Centre for Astrophysics, Department of Physics and Astronomy,
The University of Manchester, Manchester M13 9PL, UK}
\affiliation[d]{Department of Physics, Villanova University, 800 Lancaster Avenue, Villanova, PA 19085, USA}
\affiliation[e]{CNRS-UCB International Research Laboratory, Centre Pierre Bin\'etruy, 
IRL 2007, CPB-IN2P3, Berkeley, CA 94720, USA }
\affiliation[f]{Department of Physics, Harvard University, 17 Oxford Street, Cambridge, MA 02138, USA}
\affiliation[g]{Department of Astronomy, University of Maryland, College Park MD 20740 USA}
\affiliation[h]{Lawrence Berkeley National Laboratory, 1 Cyclotron Road, 
Berkeley, CA 94720, USA}
\affiliation[i]{Jet Propulsion Laboratory, California Institute of Technology, 
4800 Oak Grove Drive, Pasadena, CA 91109, USA}
\affiliation[j]{Department of Physics, Columbia University, New York, NY 10027, USA}
\affiliation[k]{Center for Computational Astrophysics, Flatiron Institute, New York, NY 10010, USA}
\affiliation[l]{Institut d'Astrophysique Spatiale, CNRS-Universit\'e Paris-Saclay, Orsay, 91405, France}
\affiliation[m]{Center for Cosmology and Particle Physics, Department of Physics, 
New York University, 726 Broadway, New York, NY, 10003, USA}
\affiliation[n]{Department of Astronomy, Columbia University, New York, NY 10027, USA}
\affiliation[o]{Kavli Institute for the Physics and Mathematics of the Universe,
UTIAS, The University of Tokyo, Kashiwa, Chiba 277-8583, Japan}
\affiliation[p]{Center for Data-Driven Discovery (CD3), Kavli IPMU (WPI),
UTIAS, The University of Tokyo, Kashiwa, Chiba 277-8583, Japan}
\affiliation[q]{Canadian Institute for Theoretical Astrophysics, 
University of Toronto, 60 St. George Street, Toronto, ON M5S 3H8, Canada}

\emailAdd{alan.j.kogut@nasa.gov}
\emailAdd{nabila.aghanim@universite-paris-saclay.fr}
\emailAdd{jens.chluba@manchester.ac.uk}
\emailAdd{david.chuss@villanova.edu}
\emailAdd{jacques.delabrouille@gmail.com}
\emailAdd{cdvorkin@g.harvard.edu}
\emailAdd{Dale.J.Fixsen@nasa.gov}
\emailAdd{shamik@lbl.gov}
\emailAdd{brandon.s.hensley@jpl.nasa.gov}
\emailAdd{jch2200@columbia.edu}
\emailAdd{bruno.maffei@u-psud.fr}
\emailAdd{anthony.pullen@nyu.edu}
\emailAdd{adityarotti@gmail.com}
\emailAdd{a.sabyr@columbia.edu}
\emailAdd{eric.r.switzer@nasa.gov}
\emailAdd{leander.thiele@ipmu.jp}
\emailAdd{Edward.J.Wollack@nasa.gov}
\emailAdd{ioana.zelko@gmail.com}

\abstract{
The Primordial Inflation Explorer (PIXIE)
is an Explorer-class mission concept
to measure
the energy spectrum and linear polarization 
of the cosmic microwave background (CMB).
A single cryogenic Fourier transform spectrometer
compares the sky to an external blackbody calibration target,
measuring the Stokes $I, Q, U$ parameters
to levels $\sim$200~Jy/sr
in each 2.65$^\circ$ diameter beam
over the full sky,
in each of 300 frequency channels
from 28~GHz to 6~THz.
With sensitivity over 1000 times greater than COBE/FIRAS,
PIXIE opens a broad discovery space
for the origin, contents, and evolution of the universe.
Measurements of small distortions from a CMB blackbody spectrum
provide a robust determination of the mean 
electron pressure and temperature in the universe
while constraining processes including
dissipation of primordial density perturbations,
black holes,
and the decay or annihilation of dark matter.
Full-sky maps of linear polarization
measure the optical depth to reionization
at nearly the cosmic variance limit
and
constrain models of primordial inflation.
Spectra with sub-percent absolute calibration
spanning microwave to far-IR wavelengths
provide a legacy data set for analyses including
line intensity mapping of extragalactic emission
and the cosmic infrared background amplitude and anisotropy.
We describe the PIXIE instrument sensitivity,
foreground subtraction,
and anticipated science return
from both the
baseline 2-year mission
and a potential extended mission.
}


\begin{document}
\maketitle
\flushbottom

% ------------------------- Main Text ------------------------- 

% \begin{spacing}{2}   		% use double spacing for submitted manuscript
\begin{spacing}{1}   		% Single spacing for drafts

% Start main text on a new page
\clearpage

\section{Introduction}
\label{sec:introduction}
The cosmic microwave background (CMB) 
records information across the history of the universe.
Its nearly-isotropic blackbody spectrum
provides compelling evidence for 
a primordial origin within a hot, dense system
in close thermal equilibrium.
The subsequent expansion and evolution 
to the presently observed structures
perturbs the CMB,
with cosmological information encoded
within its frequency spectrum,
spatial anisotropy,
and polarization.

%--------------------------------------------------------------------------
% Figure 1: CMB spectrum and residuals
%--------------------------------------------------------------------------
\begin{figure}[b]
\vspace{-5mm}			% Play with this for draft pagination
\centerline{
\includegraphics[height=4.5in]{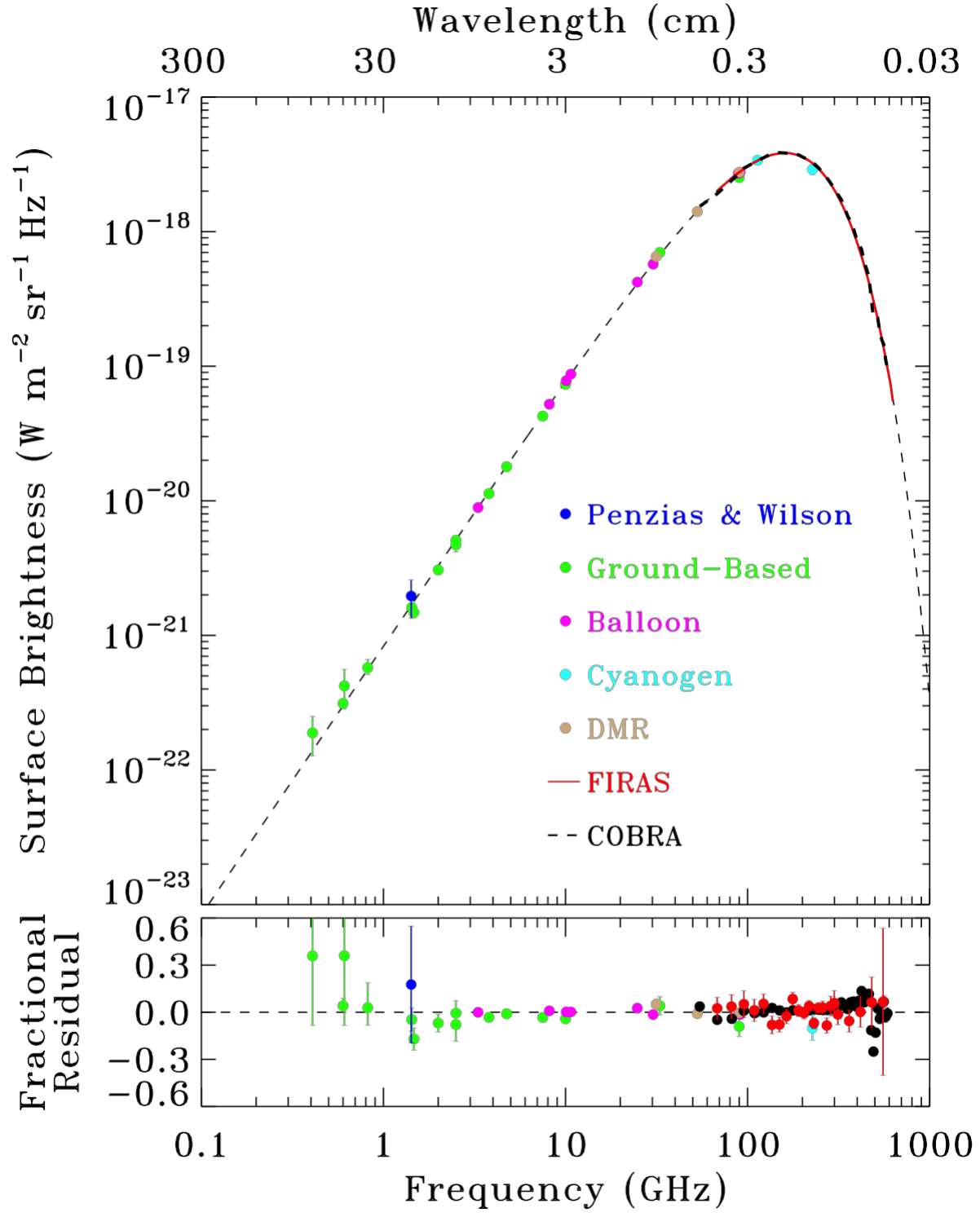}}
\caption{
Measurements of the CMB spectrum
are consistent with a single blackbody (dashed line).
The top panel shows selected precise measurements
while the bottom panel shows fractional residuals
about the best-fit blackbody.
The FIRAS residuals have been increased 1000$\times$
for visibility.
}
\label{spectrum_fig}
\end{figure}
%--------------------------------------------------------------------------

Observations of cosmological radiation backgrounds 
have played a key role 
in our understanding of the universe. 
Measurements of the CMB frequency spectrum
\cite{
penzias/wilson:1965,
Howell/Shakeshaft:1967,
deamici/etal:1985,
mandolesi/etal:1986,
kogut/etal:1988,
Bersanelli/etal:1989,
sironi/etal:1991,
deAmici/etal:1991,
levin/etal:1992,
bensadoun/etal:1993,
bersanelli/etal:1994,
staggs/etal:1996,
tris:2008,
johnson/wilkinson:1987,
staggs_balloon_1996,
arcade:2011,
meyer/jura:1985,
crane/etal:1986,
crane/etal:1989,
meyer/etal:1989,
kaiser/wright:1990,
palazzi/etal:1990,
roth/etal:1993,
dmr_syserr_1996,
firas_spectrum_1996,
cobra_spectrum_1990}
are consistent with a single blackbody
over a range of 3 decades in frequency 
and 4 decades in intensity
(Figure 1).
Data from the COBE/FIRAS spectrometer
determine the monopole temperature
$T_0 = 2.72548 \pm 0.00057$~K
\cite{fixsen:2009}.
The FIRAS limits on deviations from a blackbody spectrum 
(spectral distortions)
at levels $\Delta I/I < 5 \times 10^{-5}$ 
\cite{firas_spectrum_1996}
support the thermal hot big bang model
while ruling out alternatives such as steady state models
\cite{firas_spectrum_1994}.
Measurements of CMB anisotropies in temperature and polarization 
have provided insight 
into the contents of the universe
and their evolution from primordial density perturbations
to matter clustering,
reionization,
and the growth of large scale structure,
consistent with a single 7-parameter 
cosmological model
($\Lambda$CDM,
\cite{wmap_parameters_2013,
planck_parameters_2018}
).

Despite its success, this model is manifestly incomplete.
It requires both
dark energy and dark matter,
neither of which exist within the Standard Model of particle physics.
The observed flat geometry 
and nearly scale-invariant distribution of density perturbations
hint at an origin in a period of exponential expansion called inflation, 
but direct evidence for inflation is missing. 
The process by which baryons collapse to form galaxies 
is not fully understood, 
nor why star formation has declined by a factor of 10 
over the past 10 billion years. 

Precise measurements of the CMB 
introduce new opportunities to study the early universe  
and understand its properties.
Processes that release energy in the early universe 
can distort the CMB spectrum from a blackbody,
producing changes to both the chemical potential $\mu$ 
and inducing a characteristic Compton scattering distortion
through the Sunyaev-Zel'dovich (SZ) effect $y$ 
\cite{
zeldovich/sunyaev:1969,
sunyaev/zeldovich:1980,
danese/dezotti:1982,
bartlett/stebbins:1991,
chluba/sunyaev:2012}.
Gravitational energy released in the process of structure formation
must distort the spectrum,
while
numerous models propose other energetic processes in the early universe, 
such as primordial black holes and variants of dark matter, 
that could leave traces in the CMB spectrum 
at detectable levels
\cite{voyages_2050_wp}.
Similarly,
CMB polarization
carries information
on 
reionization and the neutrino mass (via the E-mode component)
as well as 
inflation (via B-modes).

\begin{figure}[b]
\centerline{
\includegraphics[width=6.5in]{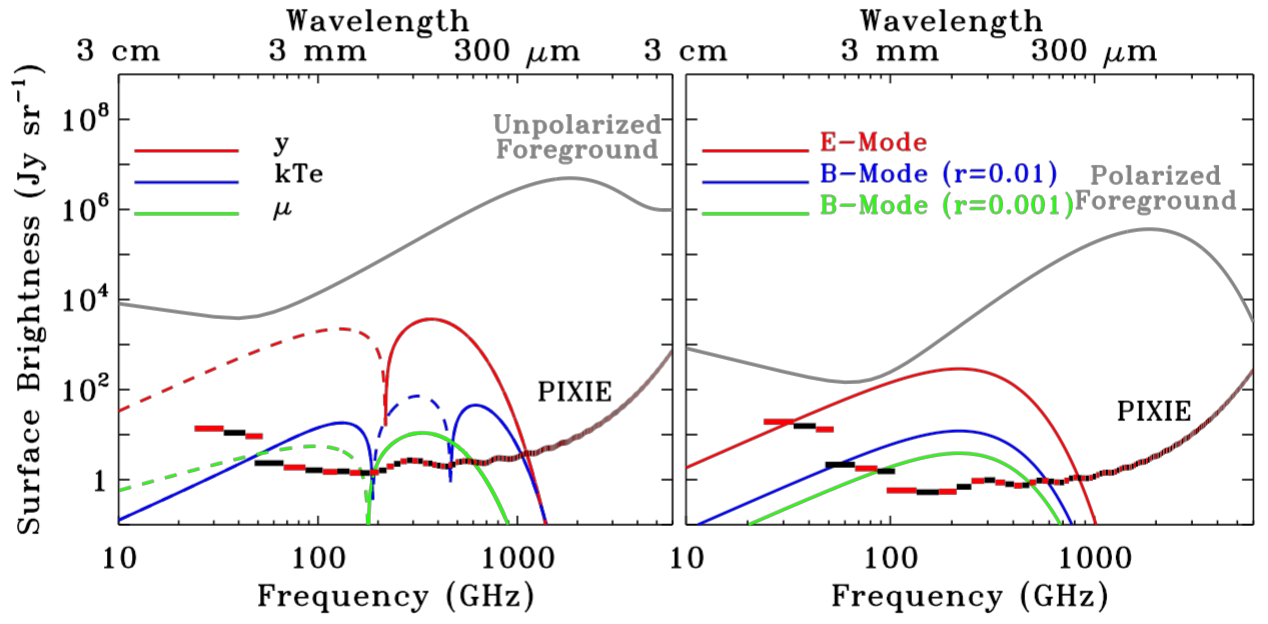}}
\caption{
Sensitivity and spectral coverage for 
the PIXIE baseline 2-year mission
in
spectral distortions (left)
and polarization (right).
Predicted cosmological signals are shown
for
the electron pressure ($y$)
and temperature ($kT_e$)
from structure formation,
the dissipation of primordial density perturbations ($\mu$),
E-mode polarization from reionization,
and
B-mode polarization from inflation
with tensor-to-scalar ratio $r=0.01$ and $r=0.001$.
Dashed lines indicate signals with negative amplitude.
}
\label{pixie_combined_sensitivity}
\end{figure}

Figure \ref{pixie_combined_sensitivity}
compares selected cosmological signals
to the median emission from astrophysical foregrounds
at high Galactic latitudes.
Cosmological signals
at amplitudes of a few Jy~sr$^{-1}$
are 2--4 orders of magnitude fainter
than the foregrounds at all frequencies.
Detecting the cosmological signals
requires both sensitivity and broad spectral coverage
to distinguish the cosmological  signals
from competing foreground emission.

%--------------------------------------------------------------------------
% Figure 2: Baseline sensitivity vs signals and foregrounds
%--------------------------------------------------------------------------
%--------------------------------------------------------------------------

The Primordial Inflation Explorer (PIXIE)
is a proposed space mission
to measure
the spectrum and polarization of the cosmic microwave background
and astrophysical foregrounds
on angular scales of 1$^\circ$ and larger
\cite{kogut/etal:2011,
pixie_calibration,
pixie_syserr:2023}.
Operating from the second Sun-Earth Lagrange point (L2),
it would survey the entire sky
in intensity and linear polarization
at integrated sensitivity of a few Jy sr$^{-1}$
in 300 spectral channels
from 28 GHz to 6 THz
(Figure \ref{pixie_combined_sensitivity}).

PIXIE was first proposed as a NASA medium-class Explorer in 2011
\cite{kogut/etal:2011}.
Previous papers have described the PIXIE
detectors
\cite{nagler/etal:2016},
optics 
\cite{pixie_feeds_2015,
pixie_feeds_2018},
calibration \cite{pixie_calibration},
and systematic error budget
\cite{nagler_syserr_2015,
pixie_syserr:2023}.
In this paper,
we review the PIXIE instrument design,
describe changes since the initial 2011 mission concept,
and
highlight the anticipated scientific results
from its baseline 2-year mission.
Section \ref{sec:pixie_mission}
describes the updated mission design,
sensitivity,
and 
foreground subtraction.
Section \ref{sec:science_goals}
reviews the principal science goals,
updates the predicted sensitivity for cosmological parameters,
and provides
new forecasts for measurements
of the cosmic infrared background,
Galactic emission lines, 
and
the star formation rate density at redshifts $z < 5$.
Section \ref{sec:discussion} 
discusses the limitations of the single-FTS instrument configuration
and
outlines the sensitivity gains possible with a larger
``Super-PIXIE'' mission.
Separate appendices
detail the data sampling and interferogram apodization,
outline the principal design features
responsible for the sensitivity improvements relative to COBE/FIRAS,
and present the full noise curves for the baseline PIXIE mission.

\section{PIXIE Mission}
\label{sec:pixie_mission}.

%--------------------------------------------------------------------------
% Figure 3: Payload and optics schematic
%--------------------------------------------------------------------------
\begin{figure*}[t]
\begin{subfigure}
  \centering
  % include first image
  \includegraphics[width=.48\linewidth]{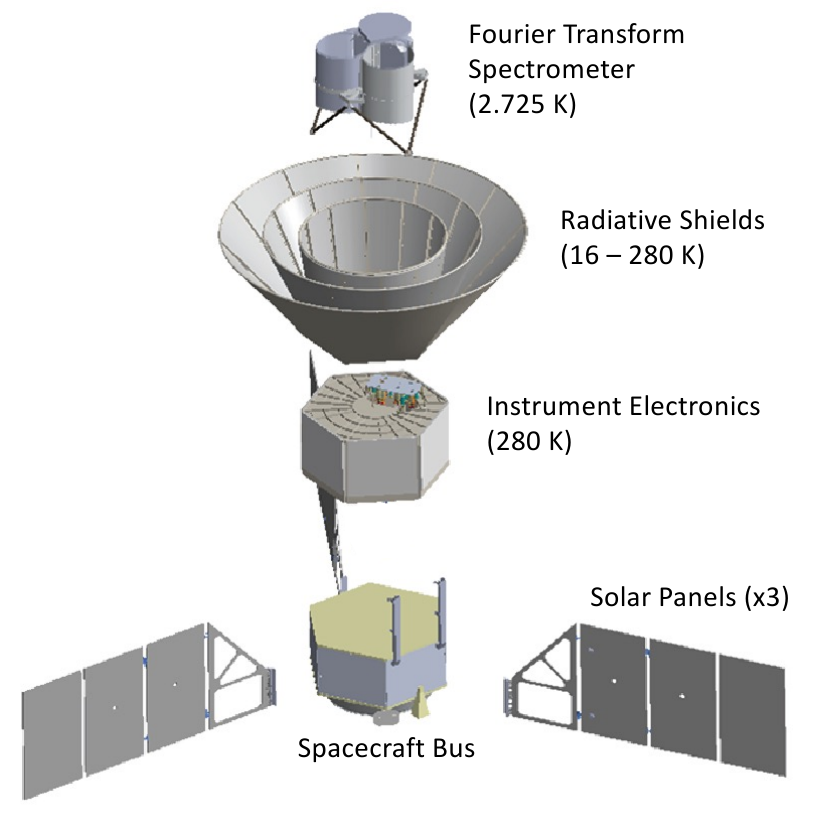}  
%  \caption{Put your sub-caption here}
%  \label{fig:sub-first}
\end{subfigure}
\begin{subfigure}
  \centering
  % include second image
  \includegraphics[width=.41\linewidth]{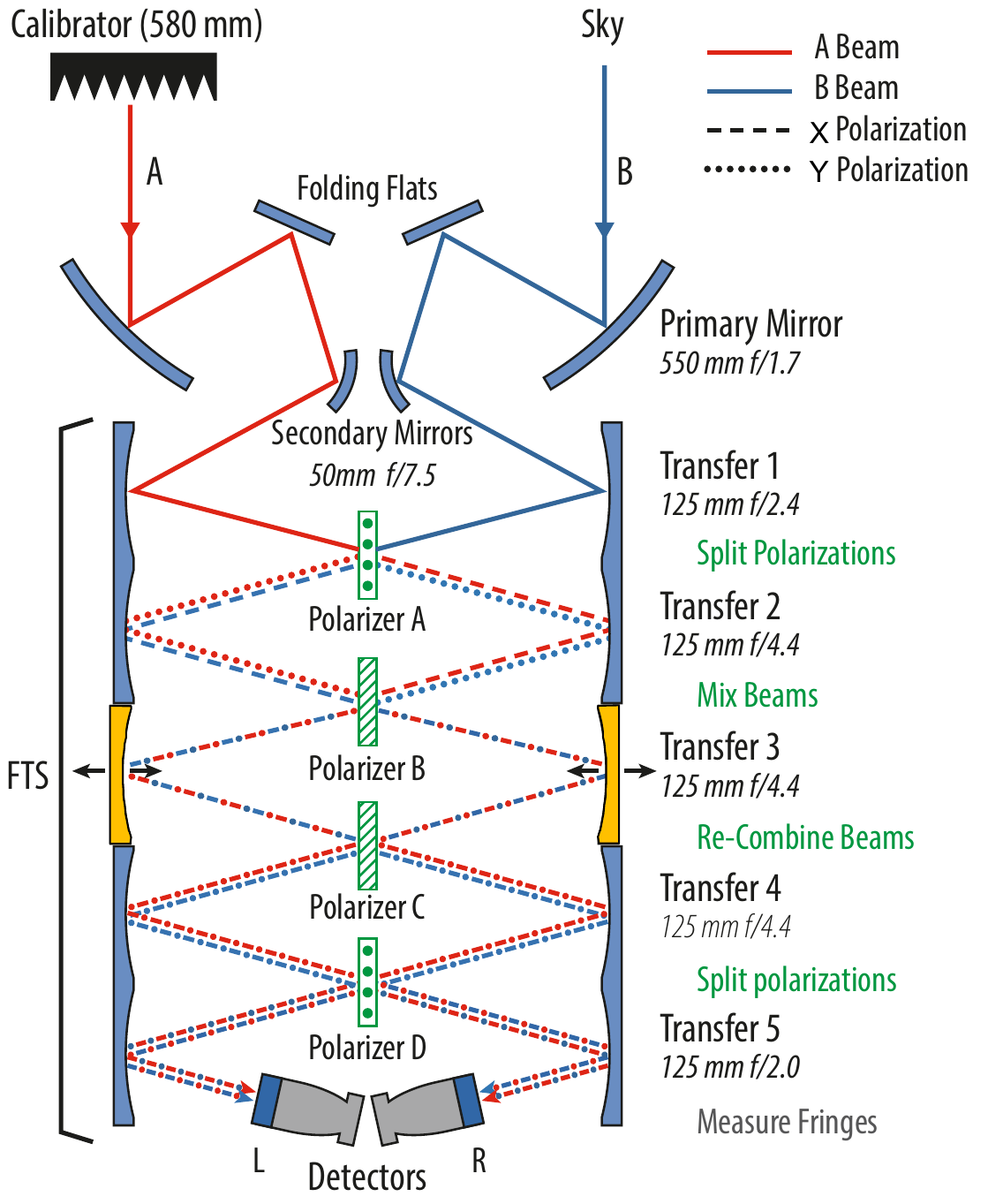}  
%  \caption{Put your sub-caption here}
%  \label{fig:sub-second}
\end{subfigure}
\caption{\label{fig:instconcept}
Left: Exploded view of the PIXIE observatory.
Nested, passively cooled shells surround 
a polarizing Fourier transform spectrometer,
which is actively cooled to $2.725$\,K to match the CMB monopole emission. 
Right: PIXIE's optical design measures both intensity and polarization 
from microwave to THz frequencies. 
As the central mirror pair moves, 
the detectors measure a fringe pattern 
proportional to the Fourier transform 
of the difference between one linear polarization from beam A 
and the orthogonal polarization from beam B.}
\end{figure*}
%--------------------------------------------------------------------------

The PIXIE instrument (Figure\,\ref{fig:instconcept})  
is a polarizing Fourier transform spectrometer (FTS) 
with two input ports and two output ports. 
The input ports are fed by two off-axis telescopes 
that produce twin beams co-pointed along the spacecraft spin axis. 
Each telescope has a primary mirror $550$\,mm in diameter. 
Folding flats and secondary mirrors route the beams to the FTS
while retaining orthogonal polarization alignment between the two beams.
Within the FTS
a set of five transfer mirror pairs, 
each imaging the previous mirror to the following one, 
shuttles the radiation through a series of polarizing wire grids. 
Polarizer\,A transmits one polarization 
and reflects the other, 
separating each beam into orthogonal polarization states
Polarizer\,B, oriented $45^\circ$ relative to Polarizer\,A, 
mixes the polarization states. 
A Mirror Transport Mechanism (MTM) moves the central mirror pair 
$\pm 4$\,mm to produce an optical phase
delay of up to $\pm 15$\,mm. 
The phase-delayed beams re-combine (interfere) at Polarizer\,C. 
Polarizer\,D (aligned to Polarizer\,A) splits the beams
again and routes them to two multi-moded concentrators. 
Each concentrator has a square aperture  
to preserve linear polarization
\cite{
pixie_feeds_2015,
pixie_feeds_2018}
and contains a pair of orthogonal polarization-sensitive bolometers 
mounted back-to-back.
The square aperture is aligned
with respect to the plane of polarization from Polarizer D
to enforce symmetry between orthogonal linear polarizations
\cite{pixie_feeds_2018}.
The four bolometers are cooled to 0.1\,K
\cite{2012Cryo...52..140S} 
to achieve background-limited sensitivity. 
The optical spectrum is the Fourier conjugate 
of the detector's time-ordered data 
across the mirror sweep. 
The mirror stroke length sets the optical frequency resolution
while the detector sampling rate determines 
the highest optical frequency. 

Each of the four detectors measures an interference fringe pattern 
(interferogram)
between orthogonal linear polarizations 
from the two co-aligned input beams. 
The spacecraft spins at $1.25$\,RPM about the beam axis 
to repeatedly interchange the $x$ and $y$ polarizations 
on the detectors. 
The Fourier transform of the observed fringe pattern 
yields difference spectra
\begin{eqnarray}
S(\nu)_{Lx} &=& \frac{1}{4} 
	\left[ ~I(\nu)_A - I(\nu)_B 
	+ Q(\nu) \cos 2\gamma + U(\nu) \sin 2\gamma ~ \right] 
	\nonumber \\
S(\nu)_{Ly} &=& \frac{1}{4}
	\left[ ~I(\nu)_A - I(\nu)_B 
	- Q(\nu) \cos 2\gamma - U(\nu) \sin 2\gamma ~\right]
	\nonumber \\
S(\nu)_{Rx} &=& \frac{1}{4} 
	\left[ ~I(\nu)_B - I(\nu)_A 
	+ Q(\nu) \cos 2\gamma + U(\nu) \sin 2\gamma ~ \right] 
	\nonumber \\
S(\nu)_{Ly} &=& \frac{1}{4}
	\left[ ~I(\nu)_B - I(\nu)_A 
	- Q(\nu) \cos 2\gamma - U(\nu) \sin 2\gamma ~\right]~,
\end{eqnarray}
where $I$, $Q$, and $U$ are the Stokes polarization parameters, 
$\gamma$ is the spacecraft spin angle, 
and $S(\nu)$ denotes the synthesized frequency spectrum 
with bins centered at frequencies $\nu$ 
set by the fringe sampling. 
$A$ and $B$ refer to the two input beams, 
while $L$ and $R$ refer to the left and right detector concentrators.

%--------------------------------------------------------------------------
% Figure 4: PIXIE scan strategy
%--------------------------------------------------------------------------
\begin{figure}[t]
\centerline{
\includegraphics[width=5.0in]{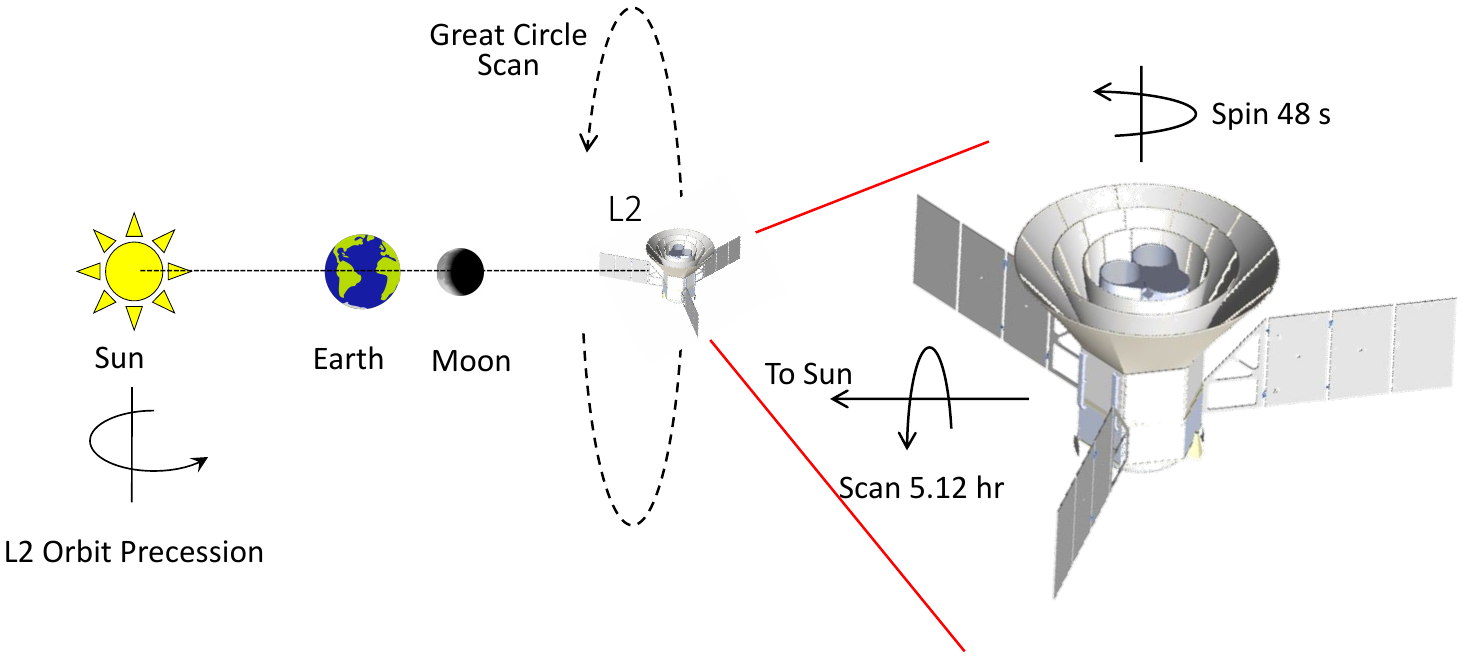}}
\caption{
PIXIE will observe at the Sun-Earth L2 point.
The spacecraft spins about the beam boresight
while simultaneously scanning the beams
in a great circle perpendicular to the sun line.
The scan pattern maps the full sky every 6 months.
}
\label{pixie_scan}
\end{figure}
%--------------------------------------------------------------------------

PIXIE carries a full-aperture blackbody calibrator 
to provide an absolute reference source
\cite{pixie_calibration}.
The calibrator can be deployed to fully cover either beam,  
or be stowed so both beams view the sky. 
When both beams view the sky, 
the instrument nulls unpolarized emission, 
so the fringe pattern encodes only the frequency spectrum of polarized emission. 
When the calibrator covers either beam, 
the fringe pattern encodes information 
for both polarization and the absolute intensity of sky emission. 
Interleaving observations with the calibrator stowed or deployed 
allows for the straightforward transfer of the absolute calibration scale 
to linear polarization while providing a valuable cross-check 
of the polarization solutions obtained in each mode.  
To control stray light, 
all internal surfaces except the active optical elements 
are coated with a microwave absorber
\cite{steelcast_2008,
chuss_rsi:2017}
forming a blackbody cavity nearly isothermal with the sky. 
Active thermal control  
maintains the telescope, FTS, and surrounding walls
within a few mK of the 2.725~K CMB temperature,
reducing thermal gradients within the instrument
to minimize effects of internal absorption or reflection
\cite{
nagler_syserr_2015,
pixie_calibration,
pixie_syserr:2023}.

PIXIE observes from a thermally-stable 
Sun-Earth Lagrange Point 2 (L2) halo orbit. 
The spacecraft spins about the instrument optical boresight at 1.25 RPM
while simultaneously scanning the boresight
through a 5.12 hour period great circle
perpendicular to the Earth-Sun line
(Figure \ref{pixie_scan}).
The annual orbital motion
precesses the great-circle orientation
to achieve full-sky coverage every six months.
The 2-year baseline mission produces 4 redundant full-sky maps
for jackknife comparison.

Data sampling is synchronous with the instrument spin,
with the mirror stroke,
spacecraft spin,
and great-circle scan
maintained in a fixed ratio
driven by a single master clock.
Several scales are relevant.
The multi-moded optics
produce a circular tophat beam\footnote{
The window function for the tophat beam
may be approximated by a Gaussian beam
with $1.65^\circ$ full width at half maximum.}
with diameter $2.65^\circ$.
The sky spectra are sorted
using the HEALPIX pixelization at resolution \texttt{NSIDE=64}
\cite{healpix/2005};
the resulting $0.9^\circ$ pixel size over-samples the beam.
The MTM produces an interferogram every 3 seconds,
the spacecraft rotates about the beam axis every 48 seconds,
and each great circle scan requires 5.12 hours (384 rotations).
During each great circle scan,
each pixel acquires data from
interferograms at 16 evenly-spaced spin angles.
Appendix \ref{sec:data_sampling} details 
the data sampling
and apodization of the sampled interferograms.

The baseline 2-year mission 
assumes that 45\% of observing time is spent with the
calibrator deployed 
(sensitive to spectral distortions and polarization)
and
45\% with the calibrator stowed
(sensitive to polarization only).
The remaining 10\% conservatively accounts for
time not used for science observations 
(MTM turnaround at the end of each stroke,
thermal settling after setpoint changes,
etc).
The calibrator deployment schedule can be altered during the mission
to provide more or less integration time in each mode.
Mission operations further assume observations 
at two different MTM stroke lengths,
providing additional sensitivity to low-frequency foregrounds.

The PIXIE instrument and mission have evolved 
since the initial 2011 concept
\cite{kogut/etal:2011}.
The 2011 version assumed observations 
from a 660 km polar sun-synchronous low-Earth orbit,
while the current version will observe from
the second Sun-Earth Lagrange point.
The L2 orbit improves thermal stability for the observatory,
reduces emission from the Earth and Moon to negligible levels,
and by decoupling the great circle scan motion
from the low-Earth orbital period
simplifies synchronization of 
the interferogram mirror stroke,
observatory spin,
and great circle scan.
Within the instrument, the optical path through the FTS has been shortened,
replacing the previous corner reflector MTM
with a pair of powered mirrors.
The 2011 concept assumed that all observations 
would use the same MTM stroke, 
and thus the same set of frequency channels.
To obtain additional low-frequency channels,
the current design assumes a combination of observations
with short and long MTM strokes
($\S$\ref{sec:sensitivity}).
The resulting sensitivity curves 
correct a factor-of-two error from the 2011 estimate
(Eq. \ref{sampled_noise_eq})
and 
now include the frequency-dependent effects
of the reflective backshort in the detector integrating cavity.
The baseline mission has been shortened from 4 years to 2 years,
based on the observing time
required for a definitive measurement of
the cosmological distortions from Compton scattering.

\subsection{Sensitivity}
\label{sec:sensitivity}

The noise equivalent power (NEP) 
from photon arrival statistics in a single linear polarization
at the detector
is given by
\begin{equation}
    {\rm NEP}^2 = 2 \frac{A \Omega}{c^2} 
    \frac{(k_B T)^5}{h^3} 
    \int \frac{x^4 dx}{e^x - 1} 
    \left ( 1 + \frac{\alpha \epsilon f}{e^x - 1} \right ) 
    (\alpha \epsilon f)
\label{mather_NEP}
\end{equation}
where 
$A$ is the detector area,
$\Omega$ is the detector solid angle,
$\nu$ is the optical observing frequency,
$T$ is the physical temperature of the source,
$\epsilon$ is the emissivity of the source,
$h$ is the Planck constant,
$k_B$ is the Boltzmann constant,
$c$ is the speed of light,
$x=h\nu/k_B T$ is the dimensionless frequency,
$\alpha$ is detector absorptivity,
and
$f$ is the power transmission through the optics to the detector
\cite{mather:1982}.
The prefactor 2 corresponds to a single linear polarization.
For non-thermal sources,
it is convenient to express the NEP in terms
of the optical power $P_\nu$,
\begin{equation}
{\rm NEP}^2 = 
	\int d\nu \left (\alpha f h \nu P_\nu  
	+ \frac{\alpha^2 f^2 P_\nu^2}{N_m} \right ) ~,
\label{power_NEP}
\end{equation}
where
$N_m =  A \Omega / \lambda^2$ 
is the number of electromagnetic modes accepted by the detector
\cite{power_NEP}.
Since the detected signal varies linearly with etendue $A\Omega$
while the noise varies as $\sqrt{A\Omega}$,
the signal-to-noise ratio
increases as the square root of the etendue.
The PIXIE detectors
each have active absorber area 1.69~cm$^2$
and etendue 4~cm$^2$~sr.
The detector absorbing structure is degeneratively doped 
to sheet resistance $377~\Omega / \Box$,
matching the impedance of free space.
The detectors are mounted within an integrating cavity
with a reflective backshort
0.689~mm behind the detectors.
The backshort increases the detection efficiency
for the CMB component
to $\alpha=0.57$,
while the efficiency
at frequencies much higher than the CMB peak
averages $\alpha=0.52$.

%--------------------------------------------------------------------------
% Figure 5: Noise Equivalent Power
%--------------------------------------------------------------------------
%
\begin{figure}[b]
\centerline{
\includegraphics[height=4.0in]{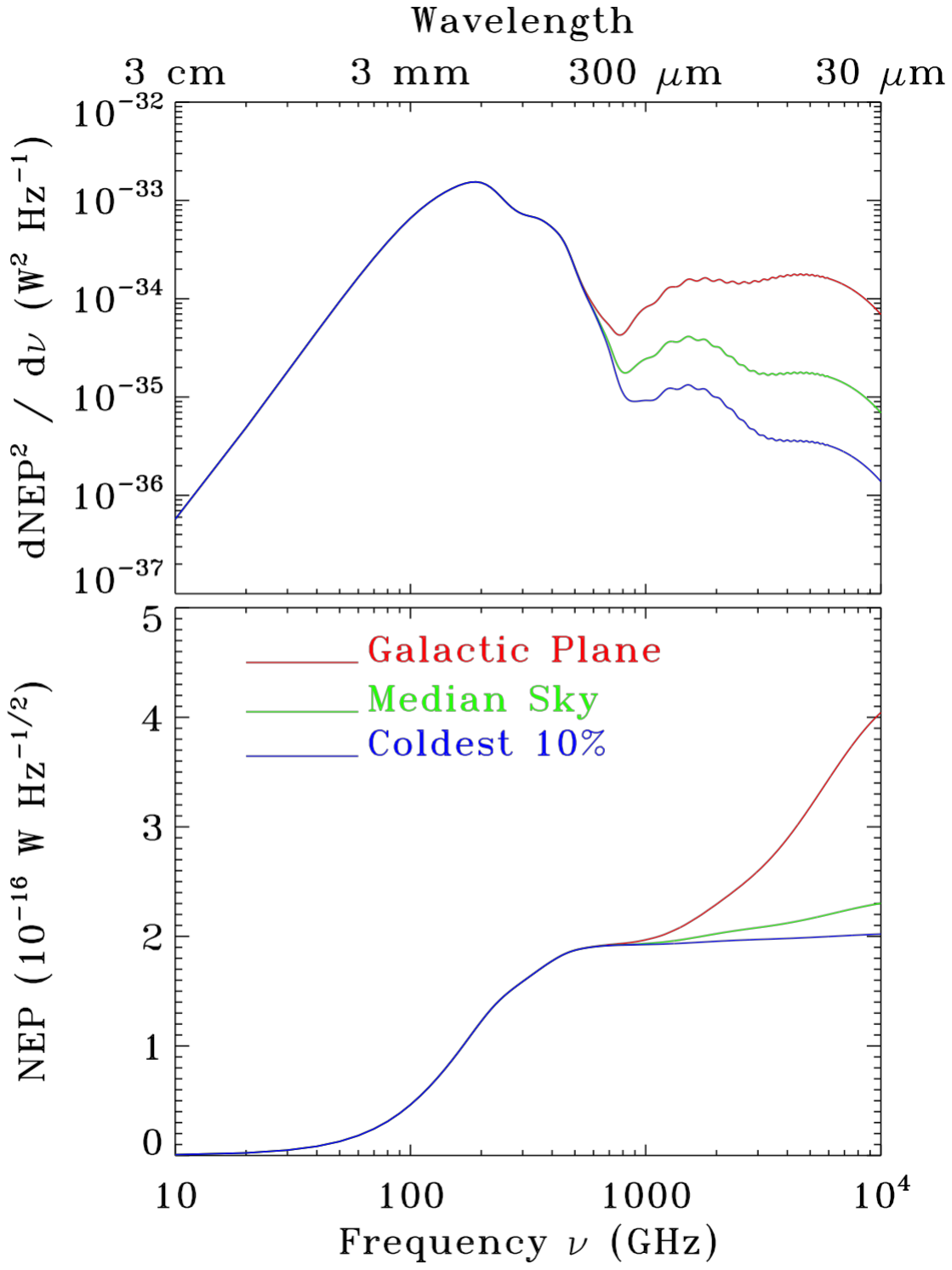}}
\caption{The photon noise equivalent power 
through the PIXIE optics 
is shown for different parts of the sky.
(top) The NEP integrand (Eq. \ref{mather_NEP}) 
is dominated by the CMB at mm wavelengths,
with minor contribution from Galactic dust,
the CIB, and zodiacal emission 
if the passband extends to shorter wavelengths.
Colors compare the photon noise 
for different lines of sight through the Galaxy.
(bottom) NEP as a function of highest optical frequency
(integral of top panel).
Scattering filters on the optics
allow the optical passband to extend to 6~THz
without significant noise penalty
over most of the sky.
\label{nep_fig}}
\end{figure}
%--------------------------------------------------------------------------

The photon noise depends on the optical power incident on the detector,
which for PIXIE is dominated by the CMB monopole or the calibrator.
The FTS optics provide throughput $f=0.9$ 
from the FTS input port to the detectors
\cite{pan/etal:2019}.
The waveguide cutoff of the concentrators
limits the throughput at frequencies below 12 GHz,
while several design elements combine to limit the high-frequency response.
The 30~$\mu$m pitch of the micro-machined detector absorbing strands
acts as a single-pole low-pass filter at wavelength 60~$\mu$m.
Two roughened mirrors within each telescope
scatter short wavelengths out of the beam
and onto the blackened walls,
producing low-pass filters
with poles at wavelength 200~$\mu$m.
The wire grid polarizers become ineffective
at wavelengths comparable to wire pitch;
these act as a set of four 60~$\mu$m single-pole low-pass filters
to reduce the fringe amplitude (signal)
but do not affect the photon noise.  
Similarly, the differential path length 
for rays at the FTS mirror centers versus edges 
washes out the fringes at short wavelengths,
again reducing signal while not affecting noise.

We estimate the optical load and photon NEP at the detector
using Eq. \ref{mather_NEP},
including contributions from the CMB 2.725~K monopole,
Galactic synchrotron, free-free, and thermal dust emission,
the cosmic infrared background,
and local zodiacal emission.
The top panel of Figure \ref{nep_fig} 
shows the differential contribution to the photon NEP
as a function of optical frequency.
The bottom panel shows the integrated photon NEP
as a function of the highest optical frequency observed.
The scattering filters
limit the high-frequency contribution 
from foreground components
so that the photon NEP is dominated by the CMB monopole
or blackbody calibrator.
Thermal dust emission
is a significant noise source
only near the Galactic plane.
Extending the optical passband from 600 GHz to 6 THz
increases the photon NEP by 16\% for the median sky brightness,
and only 5\% for the darkest 10\% of the sky,
compared to the NEP from the CMB alone.
At the median sky brightness,
the photon NEP is 
$2.2 \times 10^{-16}\,{\rm W} /\sqrt{\rm Hz}$.
For the corresponding optical load of 139\,pW
and a detector substrate temperature of 100~mK, 
the PIXIE detector adds phonon NEP
$1.5 \times 10^{-16}\,{\rm W} /\sqrt{\rm Hz}$ 
in quadrature with the photon noise
for a total
${\rm NEP} = 2.7 \times 10^{-16}\,{\rm W} /\sqrt{\rm Hz}$.

The NEP and data sampling
determine the noise in the interferograms,
\begin{equation}
\delta P_k = \frac{{\rm NEP}}{ \sqrt{2 \delta t_k} }  ~,
\label{sampled_noise_eq}
\end{equation}
where
$\delta t_k$ is the integration time 
for the $k^{th}$ interferogram sample
and
the factor of 2 accounts for the conversion
between the frequency and time domains.
After Fourier transformation,
the noise in each synthesized channel is given by the Fourier sum
\begin{equation}
\delta P_\nu =  \frac{1}{N_s} \sum_k \delta P_k \exp(i 2 \pi \nu Z_k / c / N_s) ~,
\label{fft_eq}
\end{equation}
where
$N_s$ is the number of samples
and
$Z_k$ is the optical phase delay of the $k^{th}$ sample.
The Fourier-transformed noise may in turn be referred to 
the surface brightness on the sky,
\begin{equation}
\delta S(\nu) = \frac{ \delta P_\nu }
		      { N_B \, A\Omega ~\Delta \nu ~(\alpha f) } ~,
\label{S_noise}
\end{equation}
where 
$\Delta \nu$ is the 
bandwidth of the synthesized frequency channels
after the Fourier transform.
The transmission through the optics $f$
includes the frequency-dependent signal loss
from signal dispersion and the polarizer grid efficiency,
as well as the optical scattering filters.
The factor $N_B$ accounts for the number of beams on the sky:
$N_B=1$ for measurements when the calibrator blocks one beam
(spectral distortions)
while
$N_B=2$ for measurements with the calibrator stowed
so that both beams observe the sky (polarization).

The MTM moves at constant physical velocity:
for a fixed total integration time,
the integration time $\delta t_i$ for each data sample
thus scales linearly with the desired channel width,
while the noise in the synthesized spectral channels
scales as $1/\Delta \nu$
(Eqs. \ref{sampled_noise_eq}--\ref{S_noise}).
The optimum channel width depends on the sky signal.
For line emission,
it is desirable to match the channel width 
to the width of the line profile;
broader channels wash out emission from the line.
For continuum emission the opposite is true:
the channel width should be as large as possible
consistent with retaining sufficient channels
for any necessary spectral fitting.
Since the signal is coherent while the noise is not,
the signal to noise ratio 
for continuum emission
will improve as $\sqrt{\Delta \nu}$
when integrated over channels.

%--------------------------------------------------------------------------
% Figure 6: Spectral distortions vs foregrounds
%--------------------------------------------------------------------------
%
\begin{figure}[b]
\centerline{
\includegraphics[height=3.8in]{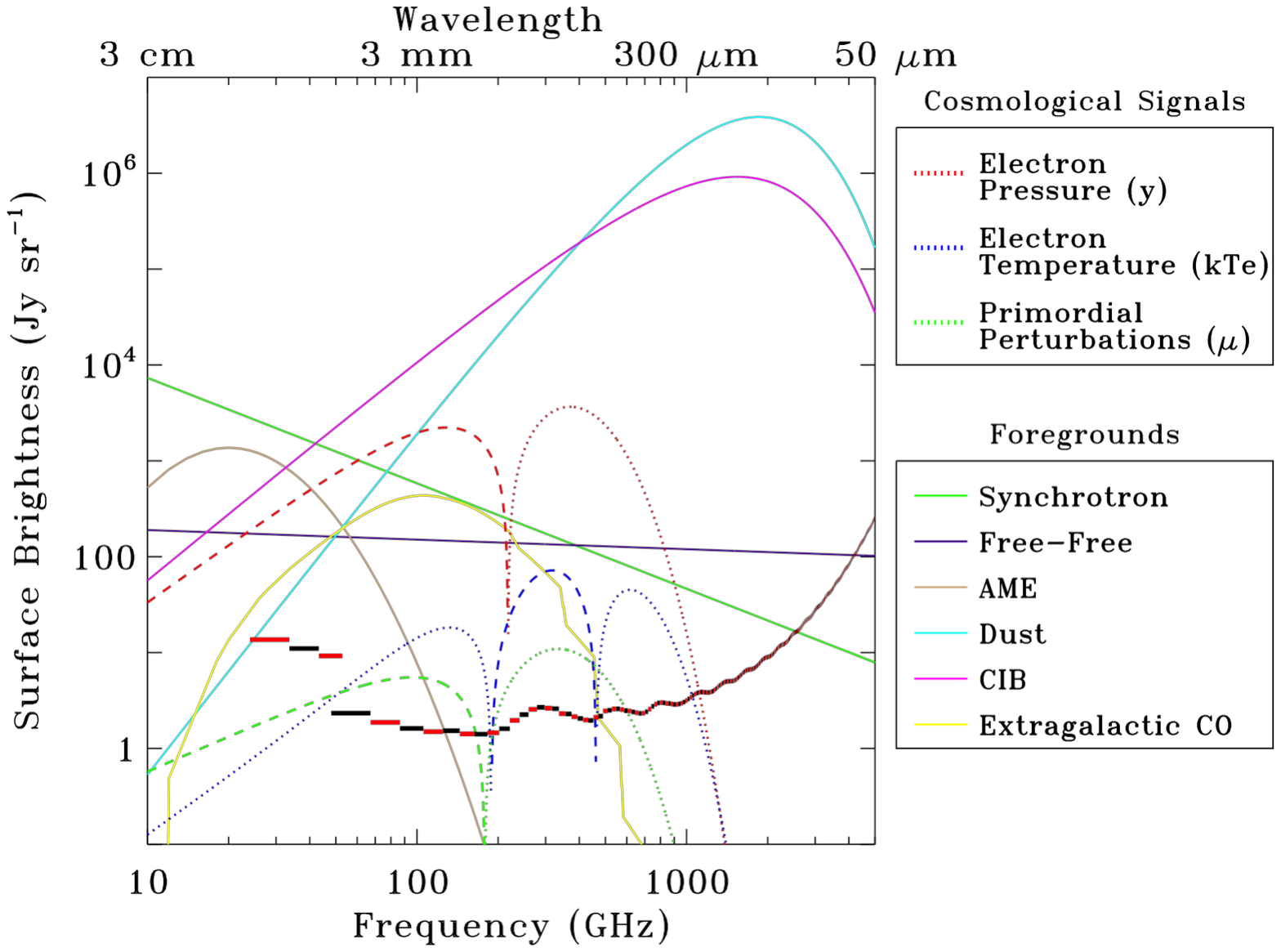}}
\caption{
Solid lines show the spectra of the principal foregrounds
at the median amplitude for Galactic latitude $|b| > 20^\circ$.
Dotted and dashed lines show the cosmological signals,
with negative amplitudes dashed.
The black/red curve shows the PIXIE noise in each spectral channel
for the baseline 2-year mission,
averaged over the cleanest 70\% of the sky.
The rise in noise at frequencies below 40~GHz
reflects the smaller fraction of observing time
spent with the MTM in long-stroke mode.
PIXIE's sensitivity to cosmological signals is limited by competing
emission from astrophysical foregrounds (see text).
\label{sd_foregrounds}}
\end{figure}
%--------------------------------------------------------------------------

The PIXIE channel width is determined by the need
to separate cosmological signals
from competing foregrounds
($\S$\ref{sec:foregrounds}).
Channel widths $\Delta \nu \sim 20$~GHz
providing spectral resolution 
$\nu / \Delta \nu \sim 15$ near the CMB peak at 270~GHz
do not provide 
sufficient resolution
to fit multiple foreground components
(synchrotron, free-free, and anomalous microwave emission)
at frequencies below 70~GH,
while
channel widths $\Delta \nu < 10$~GHz
appropriate for low-frequency foregrounds
degrade the sensitivity to CMB signals at higher frequencies.
As a compromise,
PIXIE spends 70\% of the observing time
in ``short-stroke'' mode
using broad channel widths for continuum CMB signals,
and 
30\% in ``long-stroke'' mode
with narrow widths for low-frequency foregrounds.
It is convenient to set all channel widths
commensurate with the 
CO J=1-0 line at 115.3 GHz within the Galaxy,
$\Delta \nu  = \nu_{\rm CO} / M$
so that every $M^{\rm th}$ channel
is centered on a Galactic CO line.
For measurements of spectral distortions,
PIXIE will use
$M=6$ ($\Delta \nu = 19.2$~GHz)
and
$M=12$ ($\Delta \nu = 9.6$~GHz).
The free-free and anomalous microwave foreground emission
are not polarized at relevant levels,
allowing use of wider channel widths
$M=3$ ($\Delta \nu = 38.4$~GHz)
and
$M=6$ ($\Delta \nu = 19.2$~GHz)
for polarization measurements.
Appendix \ref{sec:noise_curves}
lists the resulting channel sensitivities.

\subsection{Foreground Subtraction}
\label{sec:foregrounds}

Foreground emission from both galactic and extragalactic sources
degrades PIXIE's sensitivity to cosmological signals.
Figure \ref{sd_foregrounds} compares the principal foregrounds
to the spectral distortions
from structure formation
and the dissipation of primordial density perturbations.
Following \cite{abitbol/etal:2017},
we assess the impact of foreground emission
using the mission noise curves from this paper
in a parametric fit to the cosmological signals
and foreground emission.
The 11-parameter foreground model treats thermal dust emission 
and the cosmic infrared background
as separate modified blackbodies,
$I_\nu \propto B_\nu(T) \nu^\beta$
and fits the amplitude, temperature, and spectral index
of each component.
We model synchrotron as a power-law
$I_\nu \propto \nu^{\beta}$
and fit the amplitude and spectral index.
We fix the spectra of the remaining foregrounds
(free-free, anomalous microwave emission, and integrated extragalactic CO)
and simply fit the amplitude for each component.
An additional 4 parameters
fit the CMB monopole temperature $T_0$,
the electron pressure $y$,
electron temperature $kT_e$,
and the chemical potential $\mu$.
At frequencies above $\sim$3~THz,
the spectral energy distribution of the dust and CIB 
become more complicated;
we thus report spectral distortion limits
using only freqeuncies below 3~THz
(although higher frequencies may inform the foreground fitting
and are useful for other science goals).

Table \ref{snr_table_sd}
shows the signal to noise ratio
for the cosmological signals
estimated using a Fisher analysis
of the noise levels
integrated over the cleanest 70\% of the sky.
We adopt fiducial values
$y = 1.77 \times 10^{-6}$
and
$kT_e = 1.245$~keV
predicted for the integrated signal from
Compton scattering from groups and clusters of galaxies
\cite{hill/etal:2015}
and
$\mu = 2 \times 10^{-8}$
for the dissipation of primordial density perturbations
from a power-law distribution
consistent with the scale-invariant spectrum
seen in primary CMB anisotropies
\cite{chluba:2016}.
Results are shown for the baseline PIXIE mission
(two years of observations
with 45\% of observations devoted to  spectral distortions,
yielding 7.6 months of integration for monopole distortions
within the cleanest 70\% of the sky)
as well as an extended mission
with a total of 7 years integration time in spectral distortion mode.

% -------------- Table 1: SNR for spectral distortions  --------------
\begin{table}[t]
{
\caption{Predicted Signal To Noise Ratio for Spectral Distortions}
\label{snr_table_sd}
\begin{center}
\begin{tabular}{| c | c  | c c c | c c c |}
\hline 
Mission	&	Foreground	& $y$	& kT$_e$ & $\mu^a$ & $y$	& kT$_e$ & $\mu$	\\
\hline 
Baseline &	No foregrounds	& 4853	& 85	 & ---	 & 3101  & 80	  & 2.9 	 \\
	 &	No priors	& 103	& 6.2	 & ---	 & 18	 & 1.8    & 0.02	 \\
	 &	10\% priors	& 179	& 8.3	 & ---	 & 108   & 7.1    & 0.09	 \\
	 &	1\% priors	& 202	& 9.6	 & ---	 & 159   & 8.1    & 0.43	 \\
\hline 
Extended &	No foregrounds 	& 16\,000 & 282	 & ---	 & 10\,200  & 265    & 9.7 	 \\
	 &	No priors	& 342	& 20	 & ---	 & 61	 & 6.0    & 0.06	 \\
	 &	10\% priors	& 564	& 27	 & ---   & 213   & 19	  & 0.15	 \\
	 &	1\% priors	& 617	& 29	 & ---	 & 474   & 25	  & 0.73	 \\
\hline 
\multicolumn{8}{l}{ \footnotesize{$^a$Not fitted} } \\
\end{tabular}
\end{center}
}
\end{table}
%------------------------------------------------------------
	 	 
In the absence of foreground emission,
the baseline PIXIE mission
would detect the cosmological signals at high confidence:
3101 standard deviations for $y$,
80 standard deviations for $kT_e$,
and
2.9 standard deviations for $\mu$.
The need to identify and subtract foreground emission
degrades these values.
A minimal fit using just the PIXIE mission data
with no foreground priors
provides a robust detection
of the predicted Compton signal
(18 standard deviations for $y$ and 2 standard deviations for $kT_e$)
but falls short of a detection of the chemical potential
predicted in the standard $\Lambda$CDM model
from Silk damping of a power-law distribution of primordial density perturbations
($\S$\ref{sec:primordial_perturbations}).
We investigate the ability to improve these limits
by repeating the analysis
while placing priors on selected foreground parameters.
PIXIE data at frequencies above $\sim$600~GHz
place tight constraints on the
amplitude and spectra of the
thermal dust and CIB foregrounds
and are unlikely to be superseded by other data.
The sensitivity at lower frequencies
is limited by the fixed channel width
imposed by Fourier transform spectroscopy
($\S$\ref{sec:sensitivity})
and can be improved by including information
from measurements at lower frequencies.
We repeat the Fisher analysis
placing Gaussian priors of 10\% or 1\%
on the synchrotron and free-free parameters.
Placing 10\% priors on the low-frequency foregrounds
provides a five-fold improvement in the sensitivity to cosmological signals,
but still falls short of a detection
for the minimal $\mu$ distortion from a $\Lambda$CDM cosmology.
Even the combination of tighter (1\%) priors 
over a 7-year extended mission
produces uncertainty larger than 
the predicted chemical potential distortion.
Obtaining additional sensitivity to low-frequency foregrounds
would require
additional instrumentation 
to reduce photon noise at frequencies below $\sim$150~GHz,
extend measurements to frequencies below 30~GHz,
or both.
$\S$\ref{sec:discussion} discusses possible enhancements 
for such a ``Super-PIXIE'' design.

Covariance between the $y$ and $\mu$ distortions
affects the uncertainties for both parameters.
Table \ref{snr_table_sd} lists the predicted signal to noise ratios
with and without fitting for $\mu$.
Absent a significant detection for $\mu$,
a parametric fit excluding the $\mu$ distortion
improves the uncertainties for both
the electron pressure $y$ and temperature $kT_e$,
particularly for the case with no foreground priors.

%--------------------------------------------------------------------------
% Figure 7: Foreground decomposition for B-modes
%--------------------------------------------------------------------------
%
\begin{figure}[b]
\centerline{
\includegraphics[height=3.7in]{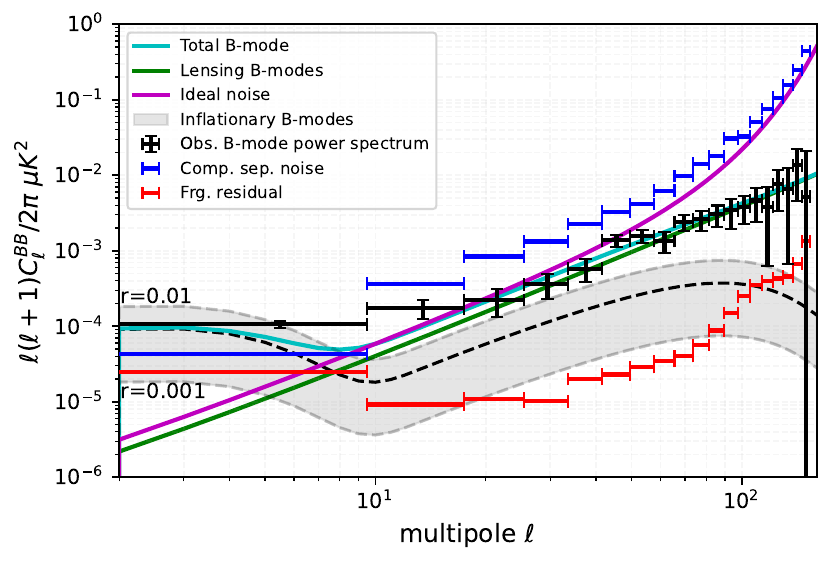}}
\caption{
Simulation results of NILC foreground decomposition 
for the PIXIE baseline mission.
The dashed line shows the $r=5 \times 10^{-3}$ inflationary signal
used for the simulation.
\label{bmode_nilc_fig}}
\end{figure}
%--------------------------------------------------------------------------

Foreground subtraction in polarization is less complex,
as only the synchrotron and thermal dust foregrounds
are expected to have significant polarization.
Figure \ref{bmode_nilc_fig} shows PIXIE's polarization capability
using a Needlet Internal Linear Combination (NILC) 
component separation
\cite{delabrouille/etal:2009,
delabrouille/cardoso:2009,
basak/delabrouille:2012,
planck_iv_2020}).
We generate a noiseless full-sky foreground model
using the Python Sky Model d1s1 variant
\cite{pysm:2017} 
and add
cosmological signals from 
E-modes,
lensing B-modes,
and primordial B-modes at amplitude $r = 5 \times 10^{-3}$.
All signals are convolved with a 
1.6$^\circ$ full width at half maximum Gaussian beam
before adding noise within each PIXIE frequency channel.
We complement the PIXIE frequency coverage
using the Planck polarization maps,
then
evaluate the resulting sky maps
on the cleanest 70\% of the sky,
using a Fisher analysis to estimate the uncertainty
$\sigma(r)$
and
a likelihood analysis
to determine the 95\% confidence upper limit $r_{95}$
\cite{hamimeche/lewis:2008}.
As with spectral distortions,
the foreground subtraction increases the effective noise
with noise amplitude (after component separation)
worse by a factor 1.8 in amplitude or 3.2 in power.
Foreground residuals are sub-dominant to the noise
and create negligible bias.

% -------------- Table 2: B-mode limits from NILC  --------------
\begin{table}[t]
{
\caption{PIXIE B-Mode Polarization Limits}
\label{nilc_pol_table}
\begin{center}
\begin{tabular}{c c c c c}
\hline 
Mission	&	$A_{\rm lens}$	& $\sigma(r)$ 	& FG Bias & $r_{95}$	\\
	&			& $\times 10^3$	& $\times 10^3$ & $\times 10^3$ \\
\hline 
Baseline &	0.3		& 0.89	& 0.39	& 2.54	\\
	 &	0.5		& 0.91	& 0.39	& 2.63	\\
	 &	1.0		& 0.98	& 0.41	& 2.86	\\
\hline 
Extended &	0.3		& 0.30	& 0.06	& 0.75	\\
	 &	0.5		& 0.33	& 0.07 	& 0.85	\\
	 &	1.0		& 0.40	& 0.08	& 1.07	\\	
\hline 
\end{tabular}
\end{center}
}
\end{table}
%------------------------------------------------------------

Gravitational lensing of the dominant E-mode polarization
creates an additional cosmological foreground
for measurements of B-mode polarization
from primordial tensor perturbations.
The power spectrum for PIXIE B-mode noise
is close to the amplitude for the un-corrected lensing foreground;
roughly 70\% of the weight for B-mode polarization analysis
results from multipoles $\ell \le 10$
where the primordial B-mode signal pulls away from the lensing foreground.
The PIXIE 1.6$^\circ$ angular resolution
does not permit de-lensing using only PIXIE data;
any de-lensing must use external data
(e.g. from CIB anisotropy or other CMB missions).
Table \ref{nilc_pol_table} shows PIXIE B-mode limits
for different levels of de-lensing,
where
the parameter $A_{\rm lens}$
shows the amplitude of the remaining (un-corrected) lensing signal.
If no de-lensing is performed
($A_{\rm lens} = 1$),
the baseline PIXIE mission
achieves
$\sigma(r) = 0.98 \times 10^{-3}$
for a 95\% confidence upper limit
$r_{95} = 2.86 \times 10^{-3}$.
Residual foregrounds
contribute a bias $r_{\rm fg} = 0.41 \times 10^{-3}$,
small compared to the noise.
Neither the B-mode uncertainty
nor the foreground  residuals
depend significantly on the assumed level of de-lensing.
The fraction of mission observations devoted to polarization
(calibrator stowed so that both beams view the sky)
can be changed throughout the mission.
An extended mission
with a total of 7 years of polarization observation
would improve the B-mode limits
to
$\sigma(r) = 0.4 \times 10^{-3}$
and
$r_{95} = 1.1 \times 10^{-3}$,
with foreground bias 
$r_{\rm fg} = 0.08 \times 10^{-3}$.
With modest delensing ($A_{\rm lens} = 0.5$),
the 95\% confidence upper limits
improve to
$ 2.6 \times 10^{-3}$ for the baseline mission 
and
$ 0.9 \times 10^{-3}$ for an extended mission. 

PIXIE will also measure E-mode polarization
to constrain the optical depth to reionization.
A similar NILC decomposition
using PIXIE and Planck data
over the cleanest 70\% of the sky
provides uncertainties
$\sigma(\tau) = 2.6 \times 10^{-3}$
for the baseline mission
and
$\sigma(\tau) = 2.5 \times 10^{-3}$
for an extended mission.
These uncertainties compare well
to the cosmic-variance limit $2.03 \times 10^{-3}$.

The above forecasts assume a simple foreground model
(d1s1) from widely used PySM sky model.
Recent Planck data suggest that variation of the
thermal dust spectral across the sky
may be larger than assumed in this model.
To investigate the impact of additional complexity in the foreground model,
we have run the NILC code on more recent models 
implemented in PySM3,
which use template maps and statistics of foreground spectra 
from Planck public data release 3
\cite{planck_iv_2020,
Planck2018_XI}.
The medium complexity model (PySM3 d10s5) 
is similar to the baseline low-complexity model
except that the dust model d10 
shows significantly more spatial variability of the dust spectrum 
in mid-to-high Galactic latitudes.
The high-complexity sky model (PySM3 d12s7a2) 
adds 2\% fractional polarization for AME, 
spectral curvature for synchrotron emission, 
and adopts a 6-layer multi-component model for polarized dust emission
\cite{3d_dust_2018}.
The medium and high complexity models
are expected to be valid 
only within the limited frequency range 1--1000~GHz.
We thus restrict the NILC analysis
to PIXIE channels below 1000~GHz,
but include an external channel with 
beam width and sensitivity comparable to WMAP K band
as an additional tracer of polarized synchrotron.

Over the cleanest 70\% of the sky,
the medium or high complexity models 
degrade the sensitivity $\sigma(r)$ by a factor 2--3
compared to the baseline model.
The reduced sensitivity is driven almost entirely 
by complexity in the dust model.
Although use of the PySM3 models
restricts analysis to frequencies below 1000~GHz,
PIXIE will produce 52 frequency channels
sensitive to polarized dust 
at frequencies 1--3.5~THz,
with median signal-to-noise ratios 25--600 
in individual high-latitude voxels within this frequency range.
Observations with the calibrator deployed
provide additional, independent data 
for both polarized and unpolarized foregrounds. 
A full treatment of foreground modeling with PIXIE
is beyond the scope of this paper
and is deferred to a future work.

\section{Science Goals}
\label{sec:science_goals}

PIXIE will survey the full sky 
at levels of a few hundred Jy per beam
for each of the Stokes I, Q, and U parameters
in 300 spectral channels
from 28~GHz to 6~THz,
reaching limits of a few Jy~sr$^{-1}$
for signals averaged over broad regions of the sky.
Table \ref{param_limits_table} lists the resulting sensitivity 
to selected cosmological parameters.
We discuss below the science goals enabled by these measurements.

\subsection{Inflation}
\label{sec:inflation}

\begin{figure}[b]
\centerline{
\includegraphics[height=3.3in]{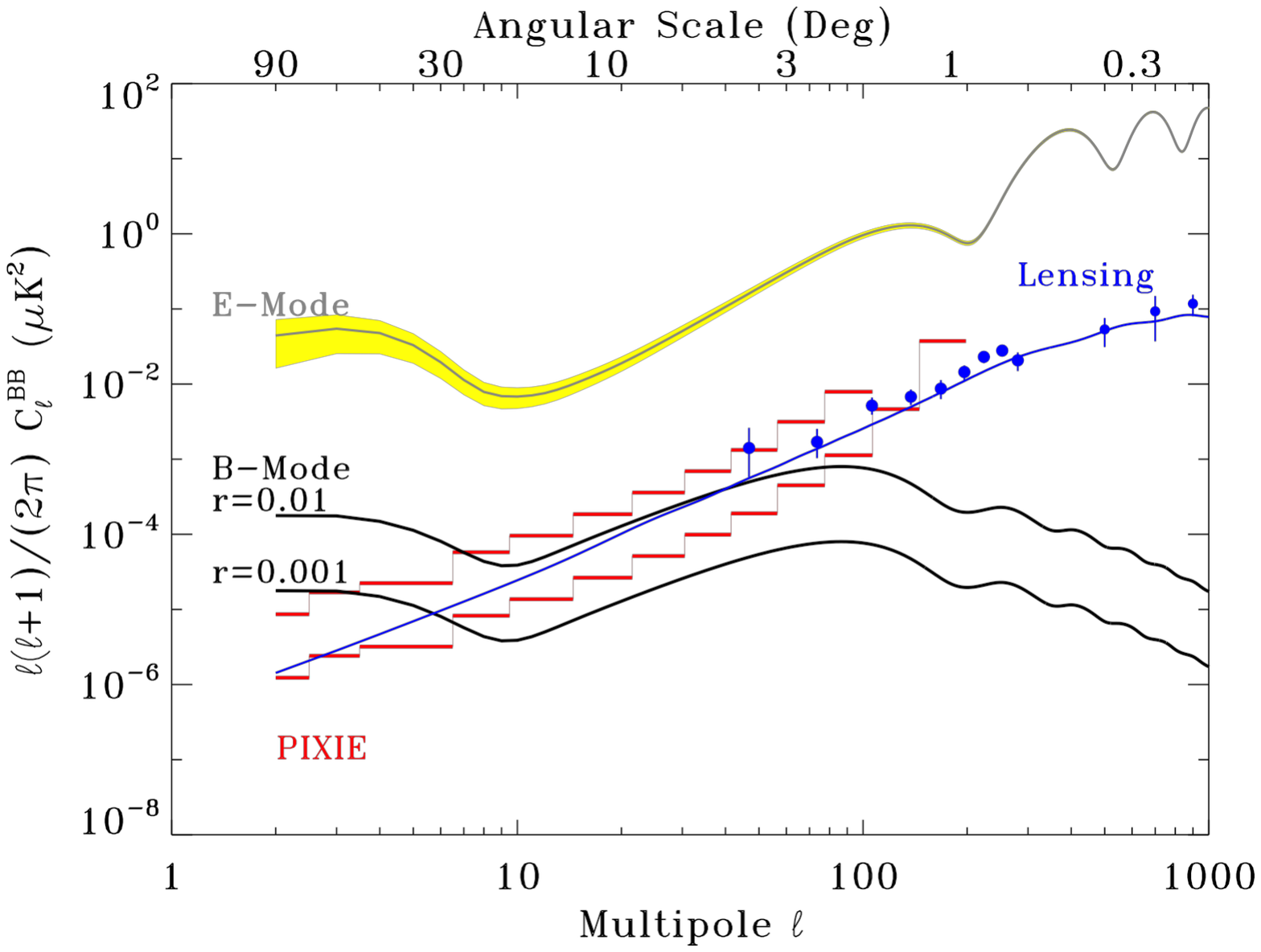}}
\caption{
Power spectra for E-mode and B-mode polarization
compared to the PIXIE sensitivity.
The shaded yellow region shows the E-mode cosmic variance.
The red lines show the PIXIE noise 
after foreground component separation 
for the baseline (upper curve)
and extended (lower curve) missions.
Blue points show current B-mode detections of the lensing foreground.
\label{pixie_cl_fig}}
\end{figure}

CMB polarization provides a test of inflation
through the B-mode gravitational wave signal 
generated during the inflationary epoch
\cite{rubakov/etal:1982,
fabbri/pollock:1983,
abbott/wise:1984,
polnarev:1985,
davis/etal:1992,
grishchuk:1993,
kamionkowski/etal:1997,
seljak/zaldarriaga:1997}.
Figure \ref{pixie_cl_fig} compares the PIXIE sensitivity
(after foreground component separation, $\S$\ref{sec:foregrounds})
to the B-mode power spectrum at values 
${r=0.01}$
predicted by the simplest models
and $r=0.001$
for a range of more complex models.
The 2-year baseline mission
spends 45\% of the total time observing polarization,
and achieves 68\% confidence uncertainty
$\sigma(r)=0.9 \times 10^{-3}$
and
95\% confidence upper limit
$r_{95} = 2.6 \times 10^{-3}$,
providing robust detection for signals at levels $r \sim 0.01$.
An extended mission
with a total of 7 years integration in polarization
improves these limits to
$\sigma(r)=0.3 \times 10^{-3}$
and
95\% confidence upper limit
$r_{95} = 0.9 \times 10^{-3}$
to constrain multi-field models.

% -------------- Table 3: Cosmological Parameter Limits  --------------
\begin{table}[t]
{
\caption{Cosmological Parameter Limits}
\label{param_limits_table}
\begin{center}
\begin{tabular}{| l | c | c | c | c |}
\hline 
Signal 			& Parameter 	& Fiducial 		& \multicolumn{2}{c|}{68\% CL Uncertainty}	\\\cline{4-5}
       			&	   	& Value    		& Baseline 		& Extended		\\
\hline	
Electron Pressure 	& $y$		& $1.77 \times 10^{-6}$ & $1.6 \times 10^{-8}$ 	& $8.3 \times 10^{-9}$	\\
\hline	
Electron Temperature 	& $kT_e$	& 1.245 keV		& 0.17 keV		& 0.065 keV		\\
\hline	
Primordial Silk Damping	& $\mu$		& $2 \times 10^{-8}$	& $2.2 \times 10^{-7}$	& $1.3 \times 10^{-7}$	\\
\hline	
Inflation		& $r$		& $10^{-3}$ --- $10^{-2}$& $0.9 \times 10^{-3}$	& $0.3 \times 10^{-3}$	\\
\hline	
Reionization		& $\tau$	& 0.05			& $2.6 \times 10^{-3}$	& $2.5 \times 10^{-3}$	\\
\hline	
Neutrino Mass		& $\sum m_\nu$	& $>$58 meV		& 13.1 meV		& 13.0 meV		\\	
\hline 
\multicolumn{5}{l}{Note: Parameter uncertainties shown after foreground marginalization assuming}  \\
\multicolumn{5}{l}{weak 10\% priors on synchrotron and free-free parameters and 50\% delensing.} \\
\end{tabular}
\end{center}
}
\end{table}
%------------------------------------------------------------

%--------------------------------------------------------------------------
% Figure 8: Polarization C_L limits
%--------------------------------------------------------------------------
%
%--------------------------------------------------------------------------

\subsection{Primordial Density Perturbations}
\label{sec:primordial_perturbations}

Primary CMB anisotropies map a nearly scale-invariant power spectrum
of primordial density perturbations
on co-moving scales $> 1$\,Mpc.
Although inflation provides a cogent explanation
for the observed large-scale structure in the early Universe,
there is as yet 
no generally-accepted model of the underlying physics
and
no compelling evidence for any specific inflationary model.
Spectral distortions provide an independent window 
to inflation and the primordial universe.
Inflation may or may not be the correct model for the primordial universe,
but primordial density perturbations exist
and source distortions from the CMB blackbody spectrum.
On co-moving scales below 1~Mpc, 
photon diffusion (Silk damping) erases the primordial fluctuations 
and transfers their energy to the CMB, 
creating a chemical potential ($\mu$-distortion) 
whose amplitude depends on the amplitude of the density perturbations 
at these scales 
\cite{daly:1991,
hu/etal:1994,
chluba/khatri/sunyaev:2012,
sunyaev/khatri:2013,
chluba:2016}.
PIXIE constraints on $\mu$ probe the amplitude of density fluctuations 
on physical scales 3-4 orders of magnitude smaller
than can be measured with primary CMB anisotropy.

%--------------------------------------------------------------------------
% Figure 9: Mu distortion limits and black holes
%--------------------------------------------------------------------------
%
\begin{figure}[b]
\centerline{
\includegraphics[height=3.7in]{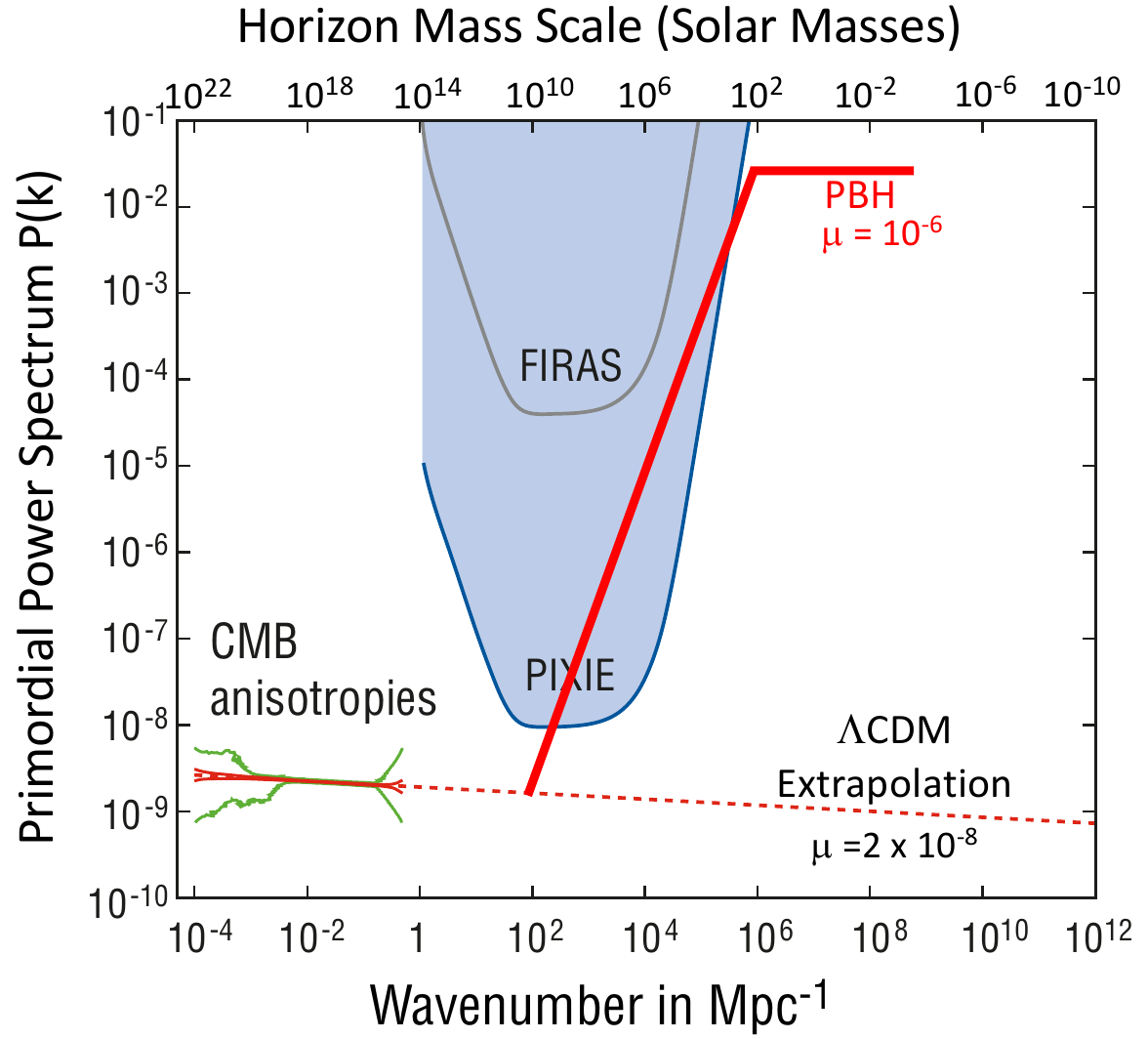}}
\caption{
Limits to the amplitude of primordial density perturbations
derived from Silk damping ($\mu$ distortions).
Dissipation of primordial density perturbations in $\Lambda$CDM predicts a      
$\mu$-distortion just below PIXIE's foreground-marginalized sensitivity.
Any detection by PIXIE requires new physics, 
such as the increased power needed to seed primordial black holes
(thick red line; see text).
\label{perturbation_limits}}
\end{figure}
%--------------------------------------------------------------------------

If the scale-invariant power spectrum 
observed in primary CMB anisotropies
persists to these smaller scales,
the standard $\Lambda$CDM cosmology
predicts spectral distortion $\mu {\sim} 2 \times 10^{-8}$
\cite{chluba:2016}.
Foreground subtraction limits PIXIE to 95\% confidence limits
$|\mu| < 9 \times 10^{-8}$
for the baseline mission
and
$|\mu| < 5 \times 10^{-8}$
for an extended mission,
assuming 1\% low-frequency foreground priors
(Table 1).
PIXIE improves the COBE/FIRAS distortion limit
$|\mu| < 9 \times 10^{-5}$
\cite{firas_spectrum_1996}
by 3 orders of magnitude,
providing new constraints on the amplitude of primordial density perturbations
at wavenumber $k = 1-10^6$~Mpc$^{-1}$
(Figure \ref{perturbation_limits}).
Upper limits at this level
constrain models of inflation,
independent of B-mode polarization
\cite{chluba/etal:2012,
baur/etal:2023,
cyr/etal:2024}.

PIXIE is not expected to detect the damping signal
from the $\Lambda$CDM cosmology;
conversely,
any positive detection
would point to physics beyond the standard model.
Primordial black holes provide an example.
Supermassive black holes (SMBHs) with mass $10^8-10^9\,M_\odot$ 
have been observed at the centers of most galaxies, 
but the detection of gravitational radiation from a population of
colliding solar-mass ($1-100\,M_\odot$) BHs  was unexpected
\cite{ligo:2016}.
Although the rate of detection 
clearly supports a current population of such BHs, 
their origin and evolution are unknown. 
They may represent a population of primordial black holes (PBHs) 
created as sufficiently-massive primordial density perturbations 
entered the casual horizon and collapsed.
No existing measurement directly constrains the primordial power 
at the relevant horizon scales.
Direct collapse of $1-100\,M_\odot$ perturbations 
requires power 
$P \sim 0.01$ 
at wavenumbers 
$10^6 < k < 10^8$\,Mpc$^{-1}$,
some 7 orders of magnitude above the level
predicted in $\Lambda$CDM
from the power observed  at scales 
$k < 1$\,Mpc$^{-1}$
by primary CMB anisotropies.
The amplitude of density perturbations
cannot jump arbitrarily,
but must smoothly transition
from the $\Lambda$CDM value 
to the much higher amplitude needed for black hole collapse.
Figure \ref{perturbation_limits} shows a simple toy model,
in which the power spectrum $P(k)$
transitions as $k^2$
from
$P \sim 10^{-9}$ at $k \sim 10^2$
to 
$P \sim 10^{-1}$ at $k \sim 10^6$
to source a population of ($1-100\,M_\odot$) BHs.
Although the perturbations directly responsible for seeding BH collapse
do not distort the CMB spectrum,
perturbations within the transition region
create a detectable signal
$\mu \sim 10^{-6}$.

Primordial perturbations will also source a stochastic background
of gravitational waves.
Enhanced power in primordial density perturbations
% at scales $10^6 < k < 10^7$~Mpc$^{-1}$
has  been suggested as a possible source
for the nHz gravitational wave background
detected by pulsar timing measurements
\cite{nanograv:2023,
epta:2023,
parkes:2023,
cpta:2023}.
Power sufficient to produce the observed pulsar timing signal
would distort the CMB spectrum
at levels up to $\mu \sim 5 \times 10^{-7}$
\cite{tagliazucchi/etal:2023,
cyr/etal:2024}.
Distortions at this level would be detectable by PIXIE.

\subsection{Particle Physics}
\label{sec:particles}

The neutrino is the only Standard Model particle
whose mass is unknown.
Measurements of E-mode polarization enable detection of the neutrino mass.
Neutrinos affect the growth of structure:
massive neutrinos cannot cluster on physical scales 
smaller than their free-streaming length,
suppressing the growth of structure on these scales.
Gravitational lensing from CMB and optical surveys 
determines the amplitude of current structure in our universe, 
but determining the growth of structure 
requires knowledge of the primordial amplitude as well.
Primary CMB anisotropy only determines the primordial amplitude
within $\sim$1\% uncertainty 
due to scattering at reionization. 
PIXIE's nearly cosmic-variance limited measurement 
of the reionization optical depth 
($\S$\ref{sec:foregrounds}) 
would remove this limiting uncertainty
to enable a determination of neutrino mass 
to 4 standard deviation precision
\cite{Calabrese:2016eii}.

Dark matter (DM) provides a more exotic test of particle physics.
The long-favored weakly interacting massive particles 
(WIMP) scenario
is seeing increased pressure 
from both
direct detection experiments
and tests for supersymmetry in particle colliders,
suggesting instead more exotic scenarios 
such as axionic DM 
\cite{
tashiro/etal:2013,
marsh:2016} 
or PBHs 
\cite{carr/etal:2010}.
PIXIE's spectral distortion limits
constrain electromagnetic interactions between
dark matter and standard-model particles
from DM decay
annihilation
and scattering
\cite{feng:2010}.
Spectral distortions from DM decay 
depend on the lifetime and abundance 
of the (excited) DM state. 
The temperature/velocity- dependence of the DM annihilation cross-section 
strongly affects annihilation signals, 
with spectral distortions being particularly valuable 
for p-wave annihilation
\cite{aalberts/etal:2018}.
Since the dark matter number density scales as 1/mass,
spectral distortions from DM annihilation or scattering 
are sensitive to low masses.
PIXIE 95\% confidence limits
$|\mu| < 5 \times 10^{-7}$
from the baseline mission
restrict Majorana annihilations
to mass $m_\chi > {\sim} 300$\,keV
\cite{mcdonald/etal:2001}.

%--------------------------------------------------------------------------
% Figure 10: Electron pressure and temperature
%--------------------------------------------------------------------------
%
\begin{figure}[t]
\centerline{
\includegraphics[height=3.7in]{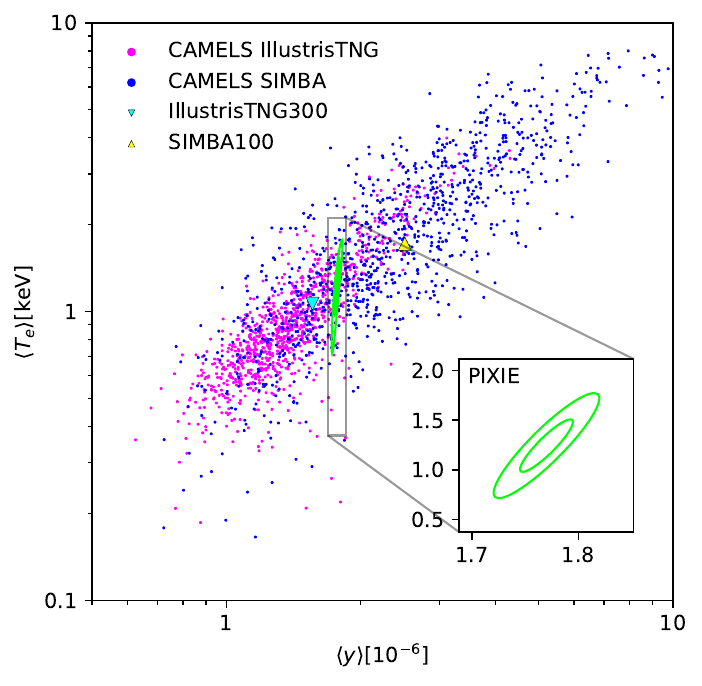}}
\caption{
PIXIE constraints on the mean electron temperature and pressure
in the post-recombination universe
compared to the current range 
used in cosmological simulations.
\label{electron_pressure_temp}}
\end{figure}
%--------------------------------------------------------------------------

\subsection{Structure Formation}
\label{sec:structure_formation}

PIXIE will determine the average pressure and temperature 
of free electrons in the post-recombination universe, 
providing critical input for numerical simulations of galaxy formation. 
Current simulations 
\cite{battaglia/etal:2010,
lebrun/etal:2014,
thiele/etal:2022}
employ a range of parameters for feedback, 
which couples the gravitational physics of large-scale structure
to the non-linear hydrodynamics on smaller (galaxy) scales. 
Energy injected into circumgalactic and intracluster gas 
from active galactic nuclei and supernovae
plays a critical but poorly-understood role in galaxy formation. 
Feedback processes shut off star formation in galaxies,
heating and expelling gas into the intergalactic medium. 
The electrons in this gas distort the CMB spectrum 
via the thermal Sunyaev-Zel'dovich (tSZ) effect.
The temperature of this gas ($10^5 - 10^6$\,K) 
is sufficiently high 
to generate non-zero relativistic corrections to the tSZ signal
\cite{hill/etal:2015,
battaglia/etal:2010}.
Assuming weak 10\% priors on low-frequency foregrounds
($\S$\ref{sec:foregrounds}),
the baseline PIXIE mission constrains
electron pressure 
$\sigma(y) = 1.6 \times 10^{-8}$
and
the electron temperature
$\sigma(kT_e) = 0.17$~keV,
providing a new window on structure formation.

Figure \ref{electron_pressure_temp} compares the PIXIE constraints 
to the range of assumptions employed in simulations. 
We adopt the baseline PIXIE mission noise curves
(Appendix \ref{sec:noise_curves})
with a total of 7.6 months integration time
on the cleanest 70\% of the sky.
In addition to the $y$-distortion and its relativistic correction, 
the Fisher forecast includes six astrophysical foregrounds
with 11 free
parameters as described in $\S$~2.2. 
Additionally, $10\%$ priors are assumed
for both the synchrotron and free-free emission, 
and no curvature parameter for the synchrotron emission 
is included in this setup.
The resulting limits differ from \cite{kogut/etal:2011}
in that both an updated noise curve and a different
baseline integration time have been used.

Measurements of monopole $y$ 
quantify the amount of feedback energy injected into the ionized gas, 
central to refining simulations.
As can be seen in Fig.~10, current models of astrophysical feedback depend
on highly uncertain parameters, 
yielding a wide range of predictions in the $y-T_e$ plane.
More precise simulations
reduce uncertainty in cosmological parameter estimates 
from large-scale structure.
Cosmological probes dependent on baryonic feedback
include weak gravitational lensing 
and the observable-mass relation of galaxy clusters.
Questions such as 
concordance between the amplitude of fluctuations 
in the early and late universe,
as well as the neutrino mass sum, 
are hindered by baryonic feedback uncertainty.
Measurements of tSZ also constrain 
primordial non-Gaussianity 
\cite{2013PhRvD..88f3526H} 
with complementarity to other probes, but at lower sensitivity.
For a fuller discussion, see
\cite{2022PhRvD.105h3505T}.

%--------------------------------------------------------------------------
% Figure 11: Star formation rate
%--------------------------------------------------------------------------
%
\begin{figure}[t]
\centerline{
\includegraphics[height=3.7in]{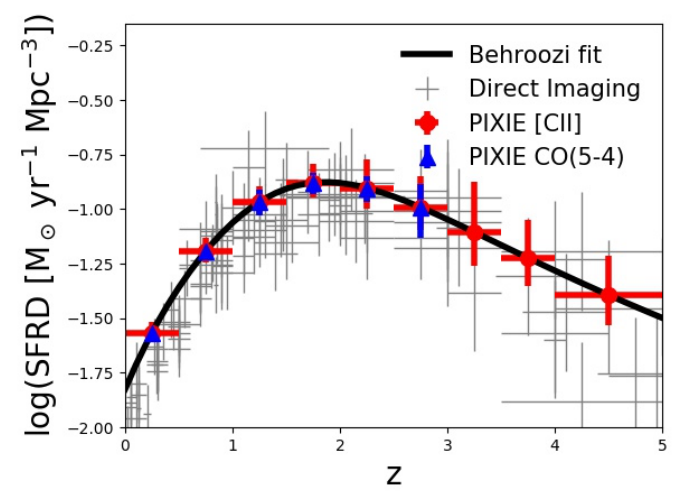}}
\caption{
PIXIE constraints on the average star formation rate density
from line intensity mapping of redshifted CO 5-4 and [CII] lines,
compared to current measurements.
PIXIE provides a legacy census of star formation across cosmic time,
unaffected by sample variance or selection effects.
\label{sfrd_fig}}
\end{figure}
%--------------------------------------------------------------------------

\subsection{Star Formation History}
\label{sec:star_formation}

Traditional spectroscopic galaxy surveys 
have been used to measure both the
star-formation-rate density (SFRD) and the metallicity. 
However, these surveys for line emission in individual galaxies 
have been limited by selection effects 
and small survey areas (sample variance), 
hampering comparison with galaxy evolution models 
\cite{2019MNRAS.482.4906P}. 
In addition, measurements of the SFRD typically rely on optical data, 
which require corrections for dust extinction
\cite{2014ARA&A..52..415M}.

PIXIE will produce an unbiased measurement 
of the star formation history  
using line intensity mapping (LIM).
LIM efficiently surveys the integral emission 
over a wide area, 
circumventing sample variance and selection effects.
PIXIE surveys a large area, 
so it is not impacted by field-to-field scatter effects. 
This is especially valuable at $z < 0.5$ 
where direct detection surveys on small areas
have high field scatter effects.
LIM with PIXIE observes the integral of all gas, 
avoiding issues relating simulations to the complex selection
\cite{2019MNRAS.482.4906P}.

Figure \ref{sfrd_fig} shows SFRD recovered from
simulations using the baseline PIXIE sensitivity.
for the  fiducial SFRD model 
\cite{2014ARA&A..52..415M}.
PIXIE will measure both [CII] and CO emission.
We assume the [CII] luminosity model from 
\cite{Padmanabhan:2019}
\begin{equation}
L_{\rm CII}(M,z) = \left(\frac{M}{M_1}\right)^{\beta} \exp(-N_1/M) 
\left(\frac{(1+z)^{2.7}}{1 + [(1+z)/2.9)]^{5.6}} \right)^{\alpha}
\label{Padmanabhan_eq}
\end{equation}
such that the [CII] luminosity is related to the SFR by a power law, 
$L \propto {\rm SFR}^\alpha$.  
The parameters $\beta$, $N_1$, and $M_1$ are set 
at all redshifts using priors from \cite{Padmanabhan:2019}.  
We set $\alpha$ at each redshift in a different manner.  
We use the Santa Cruz semi-analytic model (SAM) 
\cite{Somerville/etal:2015,
2019MNRAS.482.4906P,
Yang/etal:2022},
which predicts both [CII] luminosities and SFRs for galaxies, 
to fit $\alpha$ at each redshift using [CII] intensities and SFRDs 
from the SAM.  
For the SFRD forecast from PIXIE's CO(5-4) 
intensity measurement, we assume the Yang et al. (2019) 
CO(5-4) luminosity model, 
which is a fit to the Santa Cruz SAM.  
In this case we assume a linear relationship between the 
CO(5-4) luminosity and the SFR, 
such that the CO(5-4) intensity is proportional to the SFRD.

PIXIE will focus on CO, [CII] and [NII] emission, 
which originate from dense regions of the ISM 
and so are good tracers of both star formation
\cite{2008AJ....136.2846B,
2011ApJ...730L..13B} 
and metallicity
\cite{2012A&A...542L..34N}.
PIXIE's numerous narrow channels 
facilitate clean measurement across redshift 
\cite{2017ApJ...838...82S},
while the broad frequency coverage enables both 
[CII] and CO measurements at identical redshifts
to provide independent tests of the 
CO vs [CII] luminosity-star formation rate relations.

\subsection{Cosmic Infrared Background}
\label{sec:cib}

The Cosmic Infrared Background (CIB) 
is the integrated thermal emission
of optical and ultraviolet light
emitted by stars and active galactic nuclei,
then absorbed and re-radiated by dust.
It measures the energy release from the post-recombination universe
and
constrains the star formation history across cosmic time.
Although the CIB amplitude
is broadly consistent with known source populations,
the comparison to integrated source counts
is limited in part by the 
$\sim$6\% uncertainty in the CIB amplitude
\cite{
firas_cib_spectrum_1998,
odegard/etal:2019, 
duivenvoorden/etal:2020}.
PIXIE's absolute calibration
determines the sky brightness 
to 0.1\% accuracy
at frequencies above 600~GHz
\cite{pixie_calibration}.
Confusion with thermal emission from the diffuse Galactic cirrus
and zodiacal emission
will limit determination of the CIB monopole amplitude 
to comparable precision,
providing a target for comparison with source counts.

PIXIE will also map the CIB distribution across the sky.
The dipole anisotropy in the cosmic microwave background
is usually attributed to the motion of the Solar System
with respect to the rest frame of the
surface of last scattering at redshift $z \sim 1100$.
The observed homogeneity of the universe
on scales larger than $\sim$100~Mpc
implies that the integrated far-IR emission 
from galaxies at $z<6$
should show a comparable kinematic dipole,
with amplitude
$\Delta I/I = 1.2 \times 10^{-3}$
in the same direction
$(l,b) = (264.021^\circ, 48.253^\circ$)
as the CMB dipole
\cite{planck_2018_overview}.
Since the CIB represents the integrated emission
from all galaxies,
its dipole does not suffer from potential bias
from survey completeness.
A statistically significant difference 
in either amplitude or direction
between the CMB and CIB dipoles
would provide evidence for
a non-kinematic component for the CMB dipole
resulting from pre-inflationary physics
\cite{king/ellis:1973,
tiwari/etal:2022}.

%--------------------------------------------------------------------------
% Figure 12: Dust models
%--------------------------------------------------------------------------
%
\begin{figure}[t]
\centerline{
\includegraphics[height=3.0in]{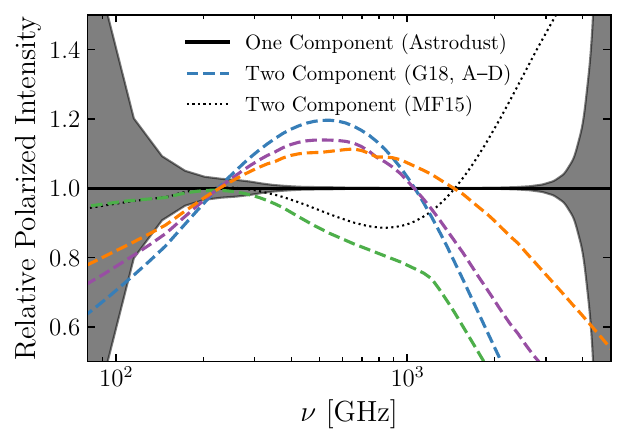}}
\caption{
Predicted dust polarization spectra 
(normalized to the astrodust model)
compared to the PIXIE sensitivity
for a single representative high-latitude line of sight.
PIXIE distinguishes between currently-viable models 
at high statistical confidence.
\label{dust_models}}
\end{figure}
%--------------------------------------------------------------------------

To quantify the PIXIE sensitivity to the CIB dipole,
we generate a sky model with the nominal
CMB, CIB, and Galactic foregrounds,
including the dominant foregrounds
from Galactic dust and zodiacal emission.
We then add simulated instrument noise
and perform a pixel-by-pixel fit of the CIB amplitude,
CMB amplitude,
and foreground amplitudes/spectral parameters
($\S$\ref{sec:foregrounds})
for 5000 realizations of the noise.
We assume that the CIB spectral energy distribution
is independent of direction on the sky,
fitting a global CIB spectrum
with amplitude fitted individually in each pixel.
Foreground parameters
(amplitude, temperatures, spectral indices)
are fit on a pixel-by-pixel basis.
The resulting distribution for the CIB in each pixel
recovers the CIB dipole
amplitude to 5\%
and direction to 4$^\circ$ uncertainty,
where the uncertainties are dominated by confusion
with the Galactic and zodiacal foregrounds.

\subsection{Interstellar Medium}
\label{sec:ism}

Interstellar dust plays critical roles 
in the chemical evolution of galaxies 
\cite{2005AIPC..761..103D,
2013EP&S...65..213A}, 
the processing of optical and ultraviolet light 
\cite{Savage1979,
Fitzpatrick1999,
Cardelli1989} 
into the infrared 
\cite{Reach1995,
Finkbeiner1999, 
Bennett2003,
PlanckCollaboration2014, 
Collaboration2016a}, 
the thermodynamics of interstellar gas 
\cite{2020MNRAS.494..146M,
2022FrASS...9.8217T}, 
and the formation of stars and planets 
\cite{Wiebe2017}.
It is a primary reservoir of interstellar metals,  
but its composition and evolution throughout the galaxy are uncertain. 
The first full-sky measurements of the polarized dust spectrum from Planck
and the unexpectedly-high fractional polarization of the dust emission
defied predictions from pre-Planck dust models 
\cite[e.g.,][]{Draine:2009}. 
These models posit that grains exist as two distinct populations 
of silicate or carbonaceous grains, 
and thus that either ISM processing is too slow to mix them 
or injection timescales are short enough 
to maintain pristine stardust populations of distinct composition. 
If so, the dust spectrum in total intensity 
should differ from that in polarized intensity
\cite{2021ApJ...909...94D}. 
This is not observed at the microwave frequencies 
accessible to Planck 
\cite{Planck2018_XI} 
or the still higher frequencies observed by the BLASTPol balloon 
\cite{Ashton:2018,
Shariff:2019}. 
Multi-component dust models have now been devised 
to respect Planck constraints 
\cite{2018A&A...610A..16G,
Ysard:2024}, 
while an alternative one-component model ("astrodust") 
has also been proposed 
\cite{2021ApJ...909...94D,
Hensley:2023}.

PIXIE's frequency coverage 
extending to the THz regime 
differentiates these dust modeling paradigms.
Figure~\ref{dust_models} 
compares models of polarized dust emission
compatible with Planck data
to the PIXIE baseline mission sensitivity.
PIXIE data at THz frequencies
discriminate between the current array of models
at tens of standard deviations
within each high-latitude line of sight.

%--------------------------------------------------------------------------
% Figure 13: Dust absorption of CMB monopole
%--------------------------------------------------------------------------
%
\begin{figure}[b]
\centerline{
\includegraphics[height=3.5in]{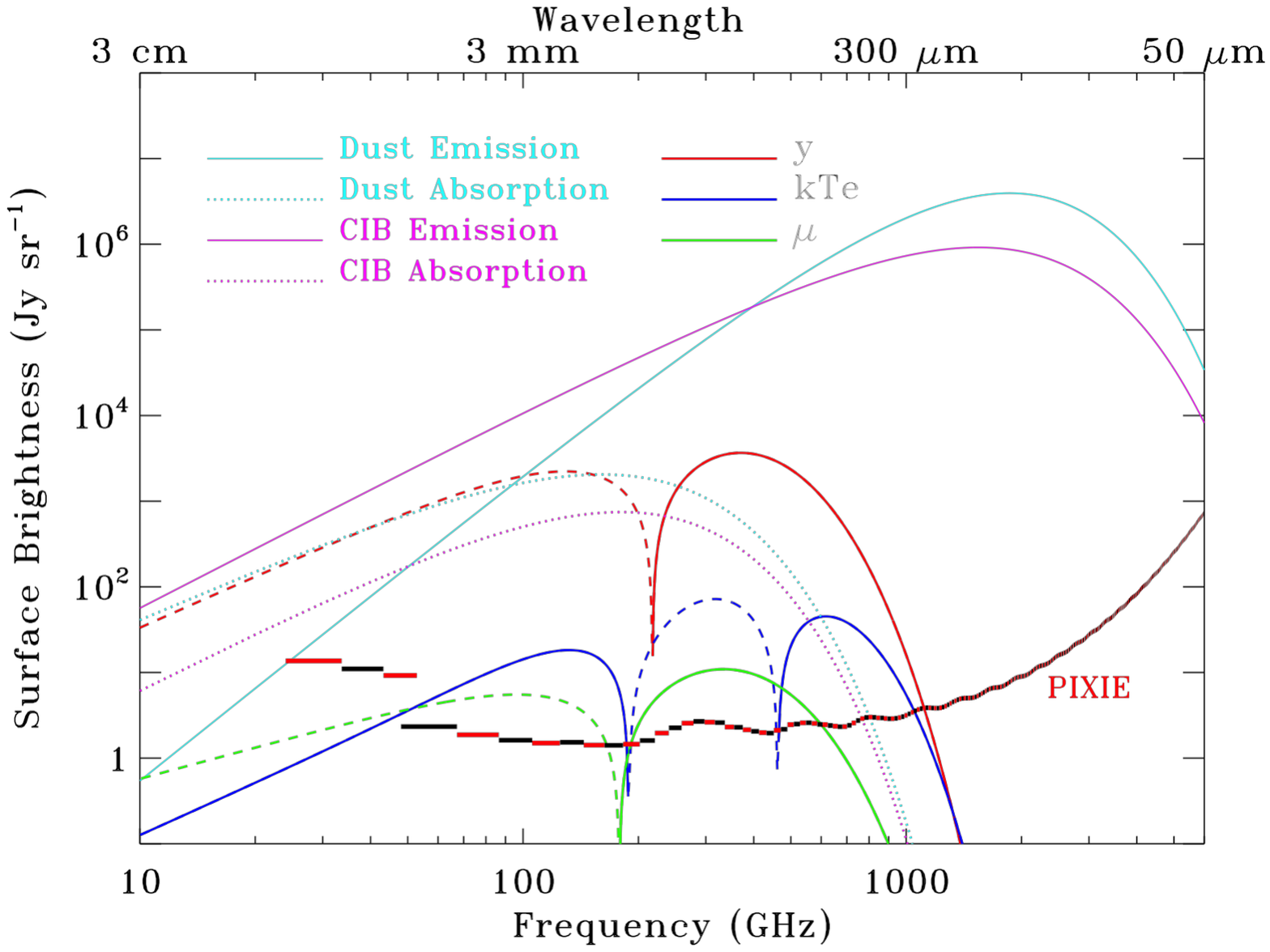}}
\caption{
Solid lines show the amplitude of emission 
from the ISM (cyan) and CIB (magenta),
while dotted lines show the absorption of the CMB monopole
by these components.
Absorption of the CMB monopole  by dust or the CIB
is comparable in amplitude
to the cosmological signals
and could be detected by PIXIE.
\label{dust_shadow}}
\end{figure}
%--------------------------------------------------------------------------

Dust absorbs as well as emits.
While correction for dust extinction is typically performed
at optical and UV wavelengths,
the low dust optical depth at much longer wavelengths
has made such corrections negligible
compared to instrument sensitivities.
Figure \ref{dust_shadow}
shows the effect of dust absorption of the CMB monopole,
sometimes called ``CMB shadow''
\cite{Nashimoto2020,
2021ApJ...914...68Z}.
The absorption signal
is comparable in amplitude
to the predicted $y$ and $\mu$ spectral distortions.
PIXIE has the sensitivity to make the first detection 
of the cosmic microwave background absorption by dust grains,
constraining the quantum behavior of dust grains 
\cite{bond1991,
wright1991,
vavrycuk2017}.

% -------------- Table 4: Line signal-to-noise ratio  --------------
\begin{table}[t]
{
\caption{Line Emission Signal To Noise Ratio}
\label{line_snr_table}
\begin{center}
\begin{tabular}{c c}
\hline 
Line	& Median SNR$^a$ \\
\hline
 CO J=1-0    &   0.1	\\
 CO J=2-1    &   2.2	\\
 CO J=3-2    &   10	\\
 CO J=4-3    &   26	\\
 CO J=5-4    &   27	\\
 CO J=6-5    &   26	\\
 CO J=7-6    &   11	\\
\hline
C+ 158 um    & 5040	\\
N+ 205 um    &  450	\\
\hline
\multicolumn{2}{l}{\footnotesize{ $^a$Median per-pixel signal to noise ratio at $|b| > 30^\circ$}} 
\end{tabular}
\end{center}
}
\end{table}
%------------------------------------------------------------

PIXIE will also map line emission from
Galactic molecular, atomic, and ionic species.
Although the moderate spectral resolution
prevents velocity-resolved measurements in the Galactic plane,
PIXIE will map the velocity-integrated line emission over the full sky.
We use low-SNR maps of the CO lines from Planck
and the singly-ionized carbon and nitrogen lines from FIRAS
to estimate the PIXIE line sensitivity at latitudes $|b| > 20^\circ$.
Table \ref{line_snr_table} 
lists the median per-pixel signal to noise ratio
for these lines.

\section{Discussion}
\label{sec:discussion}

PIXIE uses a single cryogenic Fourier transform spectrometer
to measure the Stokes $I, Q, U$ parameters
at sensitivity $\sim$200~Jy~sr$^{-1}$
in each 2.65$^\circ$ diameter beam
over the full sky,
at frequencies from 28~GHz to 6~THz.
The baseline 2-year mission spends equal time
measuring spectral distortions
(with an external blackbody calibrator deployed to block one beam)
versus
polarization
(with the calibrator stowed so both beams view the sky).
The resulting 7.6 months of integration in each mode
over the cleanest 70\% of the sky
provides a robust detection
of the spectral distortions
from the mean electron pressure and temperature
of the universe,
and provides 95\% confidence limits
to B-mode polarization
to detect the signal predicted from most single-field inflationary models.
The fraction of time spent in each observing mode
can be changed at any point throughout the mission.
PIXIE carries no expendable cryogens;
a longer mission dedicated to polarization
would improve the polarization limits 
to 
$r < 0.9 \times 10^{-3}$
to test multi-field inflation models.
Conversely, 
a longer mission dedicated primarily to spectral distortions
would lower the current 95\% confidence upper limit 
on the chemical potential
to 
$|\mu| < 10^{-7}$.
The resulting factor of 1000 improvement
over the COBE/FIRAS limits
opens an enormous discovery space
for processes including dark matter, 
primordial black holes,
and inflation.
Appendix \ref{sec:pixie_vs_firas} summarizes instrumental improvements 
compared to FIRAS.

%--------------------------------------------------------------------------
% Figure 14: Super-PIXIE comparison
%--------------------------------------------------------------------------
%
\begin{figure}[t]
\centerline{
\includegraphics[height=3.5in]{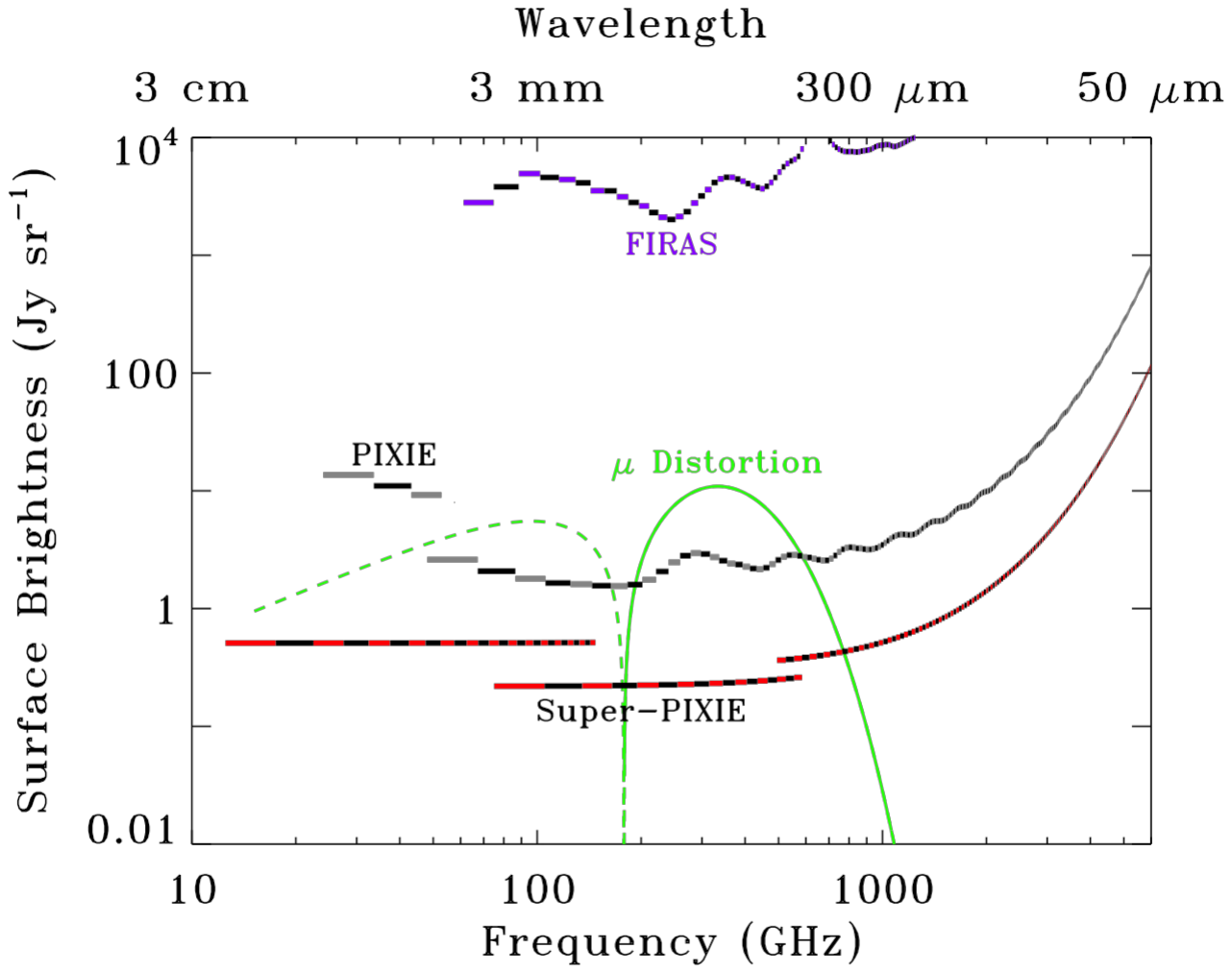}}
\caption{
A Super-PIXIE mission
with 3 FTS modules
would tune the channel width 
and optical passband of each module
to optimize the combined sensitivity to 
foregrounds and CMB spectral distortions.
\label{pixie_vs_phoenix}}
\end{figure}
%--------------------------------------------------------------------------

Foreground subtraction degrades the sensitivity
to the chemical potential $\mu$
by a factor of 6--10 compared 
to the ideal case with no foregrounds.
Improved foreground component separation
requires additional channels at lower frequencies,
without sacrificing sensitivity at higher frequencies.
This cannot be accomplished with a single FTS:
lengthening the optical phase delay to reduce the channel width
(thereby squeezing in additional low-frequency channels)
degrades the sensitivity to continuum signals
($\S$\ref{sec:sensitivity}).
As a compromise,
PIXIE spends 30\% of the integration time
in a long-stroke mode to obtain low-frequency coverage,
with the remaining 70\% spent in  short-stroke mode
to maximize continuum sensitivity.

The choice of a single FTS is dictated by the 
\$300 million (2022 USD) mission budget
for NASA's MIDEX program.\footnote{
https://explorers.larc.nasa.gov/2021APMIDEX/MIDEX/index.html}
A larger budget would allow a more ambitious instrument architecture.
This could be accomplished 
within NASA's \$1B Astrophysics Probe line\footnote{
https://explorers.larc.nasa.gov/2023APPROBE/announcements.html}
or ESA's L-class mission.
Such a ``Super-PIXIE'' 
would employ 3 (or more) cryogenic FTS modules,
each using the basic PIXIE design,
but with channel width
and optical passband (hence photon NEP)
tuned to optimize the combined sensitivity
to selected CMB signals.
Table \ref{super_pixie_table} shows one possible configuration.
A low-frequency module
would employ 5~GHz channel width
over a passband 15--150~GHz
to characterize the low-frequency foregrounds
while avoiding photon noise from the CMB monopole
at higher frequencies.
A mid-frequency module
would use wider 30~GHz channels 
over a passband 90--600~GHz
to cover the cosmological signals.
A high-frequency module
would use 30~GHz channel width
over the passband 450--6000~GHz
to characterize the high-frequency foregrounds,
setting the lowest frequency at 450~GHz
to minimize the photon noise contribution from the CMB monopole.
The two lower-frequency modules would image the sky in 7 sky pixels
to increase the system etendue.
Figure \ref{pixie_vs_phoenix} 
compares the resulting sensitivity 
to both PIXIE and FIRAS.
Such a Super-PIXIE mission
could provide a $3\sigma$ detection
of the minimal chemical potential distortion
from dissipation of primordial density perturbations
within the $\Lambda$CDM model.

% -------------- Table 5: Super-PIXIE configuration  --------------
\begin{table}[t]
{
\caption{Potential Super-PIXIE Configuration}
\label{super_pixie_table}
\begin{center}
\begin{tabular}{c c c c}
\hline 
FTS		& Channel	& Lowest 	& Highest	\\
Module		& Width	   	& Frequency	& Frequency	\\
\hline
Low-Freq	& 5 GHz		& 15 GHz	& 150 GHz	\\
Mid-Freq	& 30 GHz	& 90 GHz	& 600 GHz	\\
High-Freq	& 30 GHz	& 450 GHz	& 6 THz		\\
\hline 
\end{tabular}
\end{center}
}
\end{table}
%------------------------------------------------------------

\acknowledgments

This research was carried out in part at the 
Jet Propulsion Laboratory, California Institute of Technology, 
under a contract with the National Aeronautics and Space Administration.
The cost information contained in this document is of a 
budgetary and planning nature 
and is intended for informational purposes only. 
It does not constitute a commitment on the part of JPL and/or Caltech.

%-------------------------------------------------------------------
% Appendix A: Data sampling
%-------------------------------------------------------------------

% For now, start appendices on new page
\clearpage

\appendix
\section{Data Sampling and Apodization}
\label{sec:data_sampling}

\begin{figure}[t]
\centerline{
\includegraphics[height=2.5in]{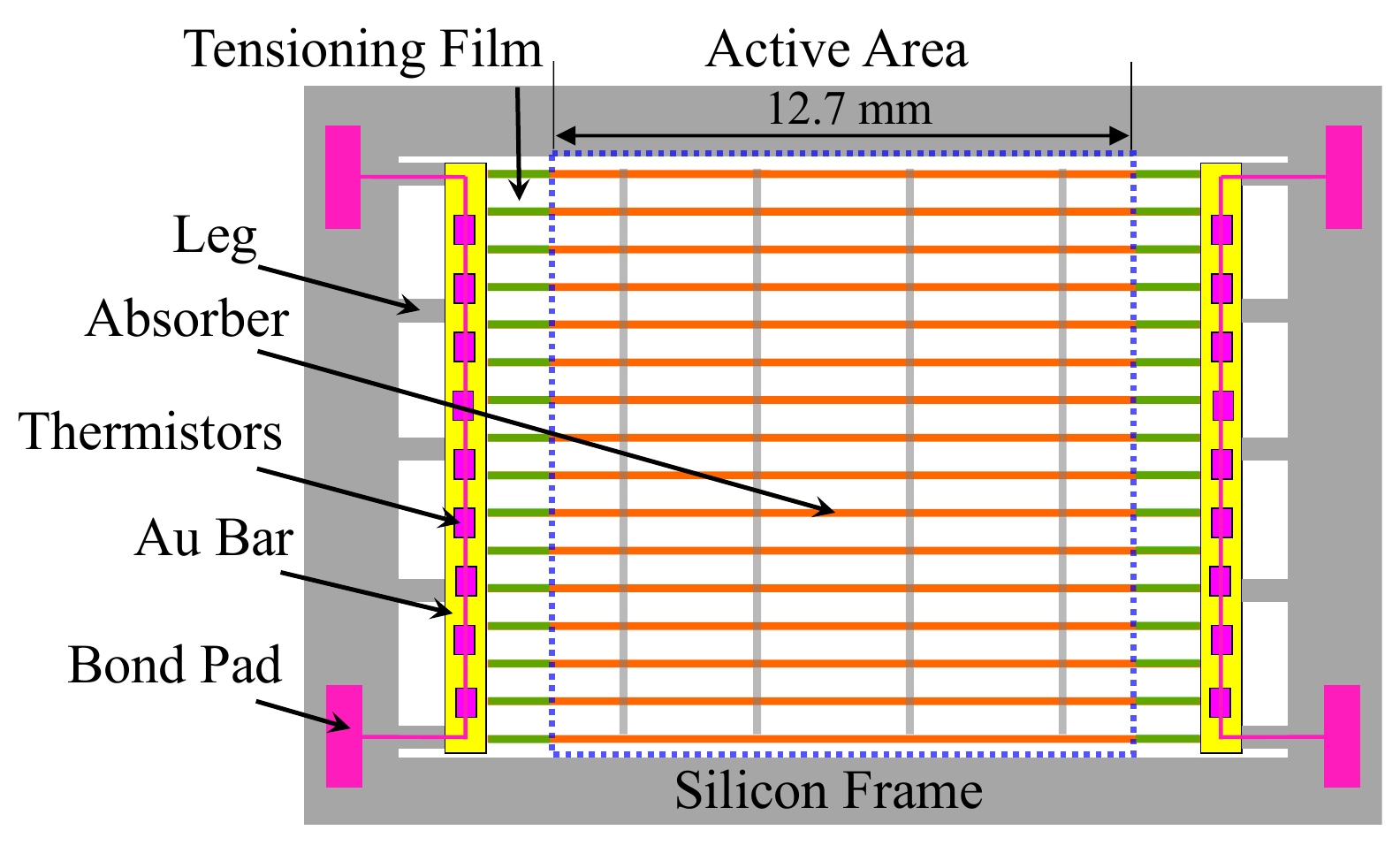}}
\caption{
Schematic showing major elements of the PIXIE detectors.
Doped silicon wires absorb a single linear polarization
over etendue 4 cm$^2$ sr.
}
\label{detector_schematic}
\end{figure}

PIXIE uses physically large, multi-moded detectors
to operate at background-limited sensitivity
across a broad spectral range.
Figure \ref{detector_schematic} shows the detector design.
The detector absorbing element consists of 3 $\mu$m wide wires of 
micromachined crystalline silicon
12.7 mm long and spaced 30 $\mu$m apart.
The absorber wires are
degenerately doped with phosphorus
to be metallic at all temperatures
and absorb a single linear polarization.
Un-doped cross-members every 300~$\mu$m
maintain alignment;
the structure is 85\% open
to minimize the cross section for cosmic ray impacts.
A tensile Al$_2$O$_3$ film
deposited on the wires outside of the active absorbing area
maintains grid planarity within $\pm 5 ~\mu$m
and raises the absorber resonant frequency above 1 kHz.
Silicon legs suspend the bolometer 
and provide thermal isolation from a silicon frame heat sunk at 100 mK.
Gold bars $0.5\,\mu {\rm m}$ thick 
at the end of the absorbing structure 
dominate the heat capacity
to stabilize the thermal response
\cite{nagler/etal:2016}.
Ion-implanted silicon thermistors measure the 
temperature fluctuations of the absorbing structure 
in response to optical power. 
Two identical arrays are hybridized back-to-back 
to give dual-polarization sensitivity
within each optical concentrator.
Tensioned leads connect each bolometer to a cryogenic JFET amplifier,
mitigating capacitive microphonic contamination of the signal band
\cite{suzaku_JFET,
hitomi_JFET}.
The amplifier is AC biased above the $1/f$ knee of the JFET.

%--------------------------------------------------------------------------
% Figure A1: Detector schematic
%--------------------------------------------------------------------------
%--------------------------------------------------------------------------

%--------------------------------------------------------------------------
% Figure A2: MTM performance
%--------------------------------------------------------------------------
\begin{figure}[t]
\vspace{-5mm}			% Play wiith this for draft pagination
\centerline{
\includegraphics[height=3.0in]{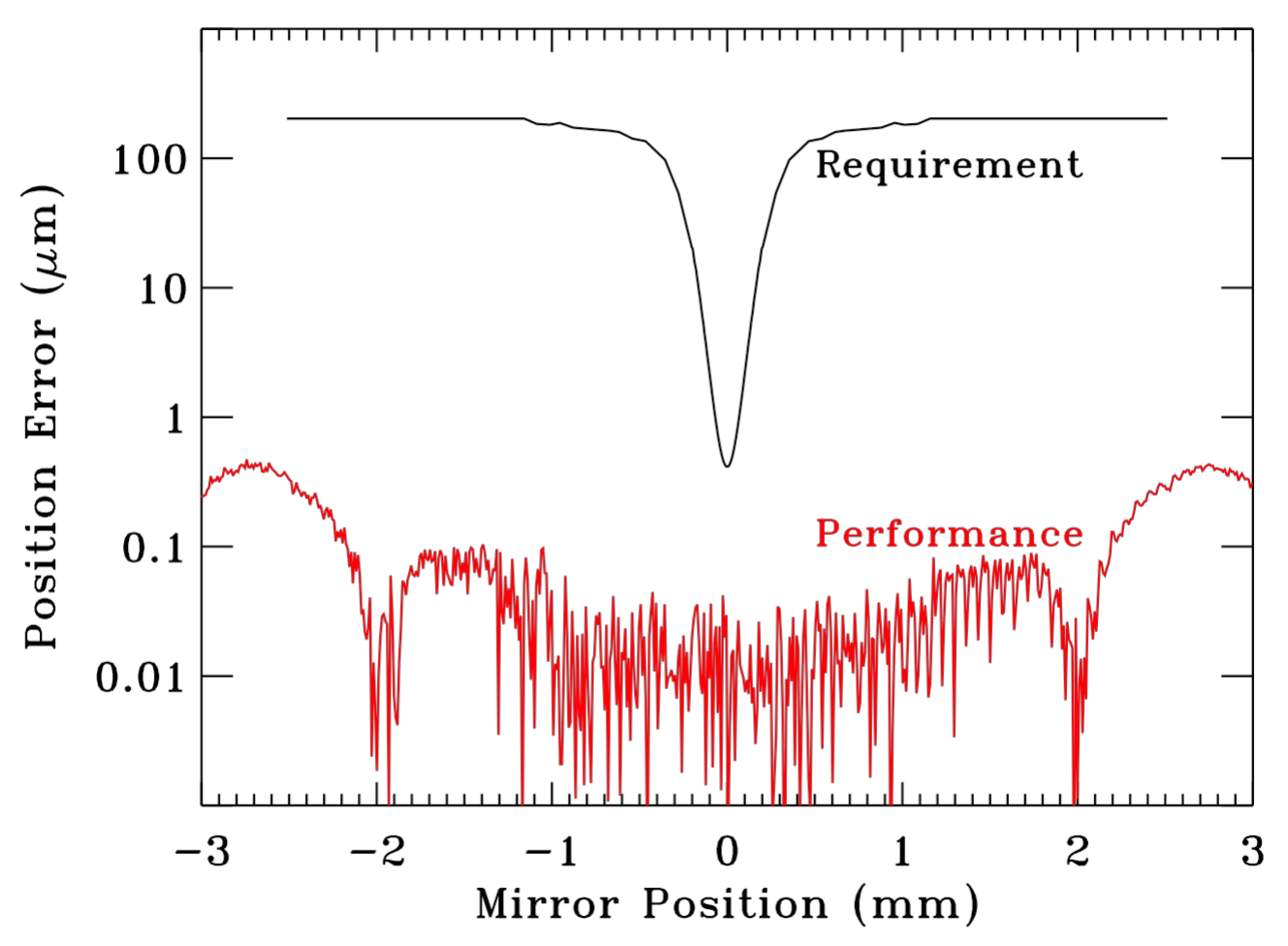}}
\caption{
The measured position jitter for an engineering prototype 
phase delay mechanism 
exceeds the PIXIE requirements over the entire optical range.
}
\label{MTM_fig}
\end{figure}
%--------------------------------------------------------------------------

A double-sided interferogram with $N_s$ samples
over optical phase delay 
$[-L, L]$
returns spectra 
in $N_s/2$ channels at center frequencies
$ [0, 1, 2, ..., N_s/2] ~ \times \Delta \nu$
where
$\Delta \nu = c/(2L)$ is the synthesized channel width.
The maximum mirror throw
determines the channel width,
while the number of samples
determines the number of channels
and hence the highest synthesized frequency.
The MTM mirror hardware 
moves at constant phase velocity $v = 4$~mm~s$^{-1}$
over maximum optical phase delay $\pm$16~mm.
The 512~Hz detector data rate
thus provides 2048 samples 
for a single end-to-end mirror stroke
over the maximum optical range,
corresponding to 1024 synthesized channels
with channel width
$\Delta \nu = 9.6$~GHz
to maximum frequency 9.8~THz.
Software commands allow shorter mirror strokes
and thus wider synthesized channels.
The constant mirror velocity
produces a fixed relation between
the audio frequencies of the sampled data
and the corresponding optical frequencies,
\begin{equation}
\nu_{\rm data} = \frac{v}{c} \nu_{\rm opt} ~.
\label{beta_eq}
\end{equation}
Scattering filters limit the optical passband
to frequencies below 6~THz,
corresponding to
audio frequencies below 80~Hz;
synthesized channels at higher optical frequency
contain noise but no optical signal.

%--------------------------------------------------------------------------
% Figure A3: Channel-to-channel covariance
%--------------------------------------------------------------------------
\begin{figure}[b]
\vspace{-5mm}			% Play with this for draft pagination
\centerline{
\includegraphics[height=3.0in]{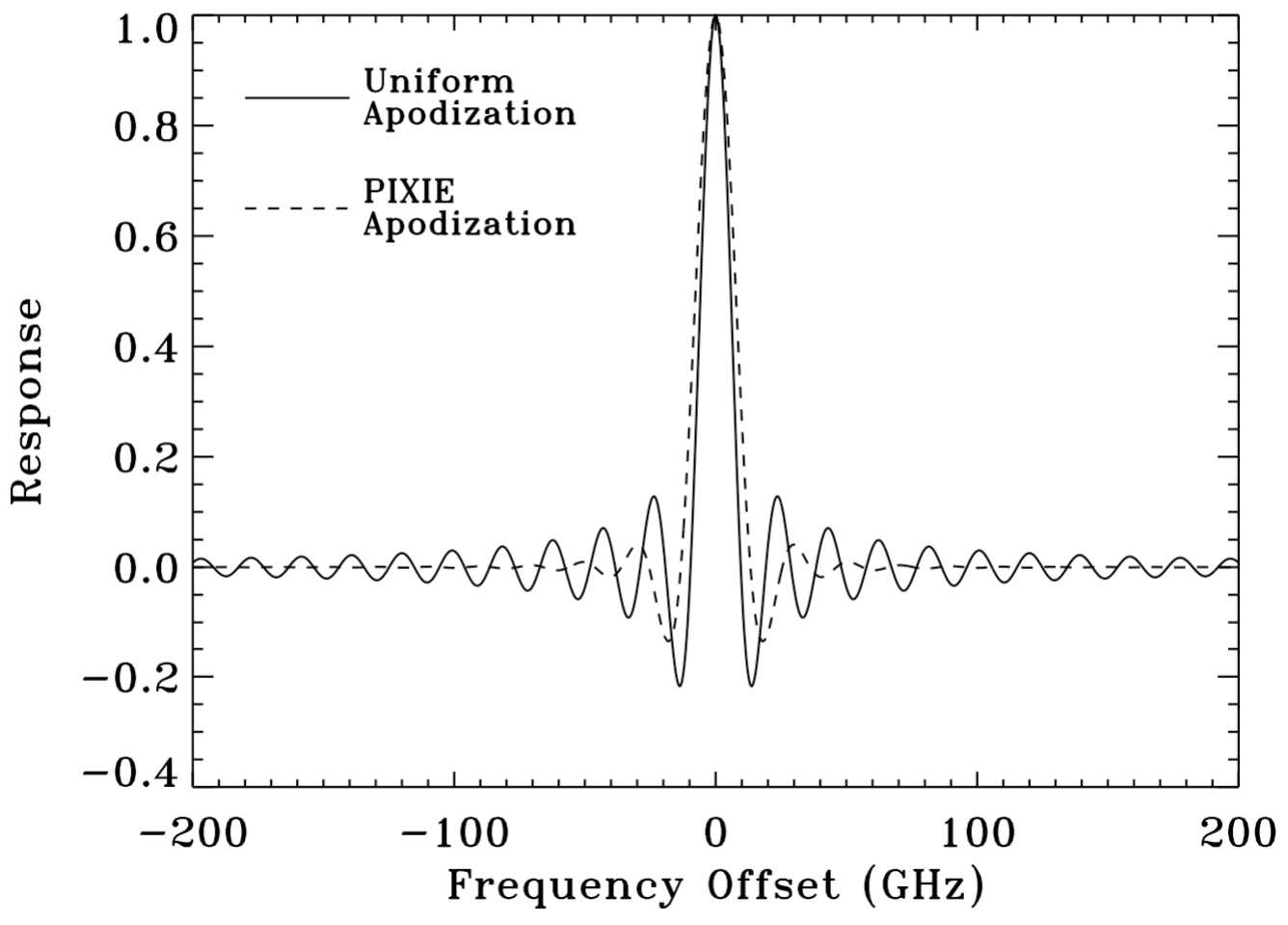}}
\caption{
Frequency response after Fourier transform
for different apodization of the sampled interferograms.
The PIXIE apodization has broader channel width
but much less ringing 
than uniform apodization.
}
\label{pixie_covar}
\end{figure}
%--------------------------------------------------------------------------

PIXIE acquires data with the mirror moving in both directions.
To minimize aliasing in the synthesized frequency spectra, 
the phase-delay mirror position 
must be controlled to $0.4 \,\mu {\rm m}$
precision near the white-light null 
at zero optical phase delay.
Figure \ref{MTM_fig}
compares the PIXIE requirement
to the performance of an engineering prototype.
The mirrors are mounted on a double flexure suspension 
with a voice coil drive motor.
A non-contacting capacitive position sensor provides position feedback 
within an active control loop. 
Passive electromagnetic dampers on both the mirror drive and flexures 
provide additional control stability.
The measured position jitter
exceeds requirements over the entire optical range.

%--------------------------------------------------------------------------
% Figure A4: MTM strokes and apodization
%--------------------------------------------------------------------------
\begin{figure}[t]
\vspace{-5mm}			% Play with this for draft pagination
\centerline{
\includegraphics[height=2.8in]{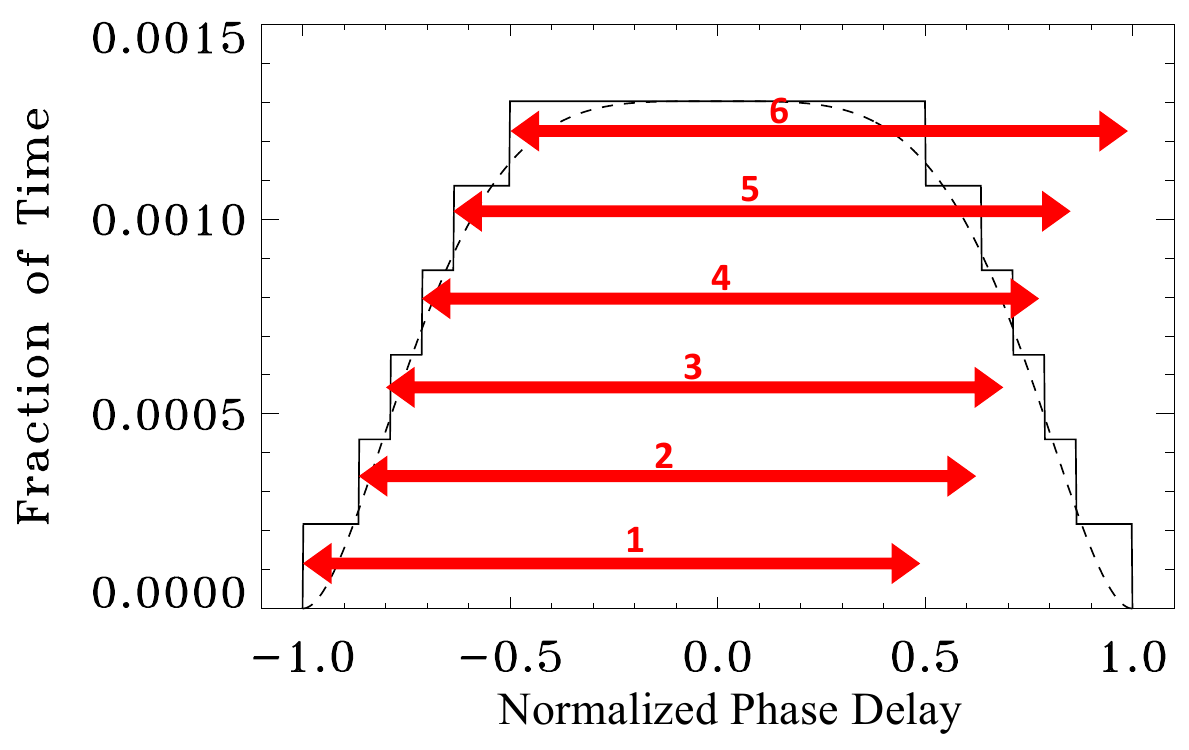}}
\caption{
PIXIE varies the MTM stroke to apodize the observed interferograms.
Each stroke uses 3 seconds to cover 3/4 of the total optical phase delay.
A set of 6 different start/stop points (red bars)
provides an integration time at each optical phase delay
(solid line)
closely approximating the desired apodization
(dashed line).
}
\label{mtm_stroke}
\end{figure}
%--------------------------------------------------------------------------

Apodization of the Fourier transform
determines the channel-to-channel covariance.
For a given mirror throw,
uniform weighting of the interferogram
produces the narrowest channel width;
however,
the resulting spectra
have channel-to-channel coupling
described by a sinc function across a broad spectral range.
PIXIE will use the apodization function
\begin{equation}
W(x) = (1 - x^4)^2
\label{apod_eq}
\end{equation}
where
$x = z/z_{\rm max}$
is the fractional phase delay.
Figure \ref{pixie_covar}
compares the frequency response
for each choice of apodization.
The PIXIE apodization 
increases the channel width by 37\% 
compared to uniform apodization,
but greatly reduces the ringing over larger frequency separations.

The apodization requires lower weight (fewer observations) 
for samples near the maximum optical phase delay.
Continuously stroking the MTM over its full range every 4 seconds
and then de-weighting samples (in the Fourier transform software)
to achieve the desired apodization
results in an effective loss of 29\% of the available observing time.
Instead, PIXIE uses a shorter MTM stroke
covering 3/4 of the total range every 3 seconds,
while periodically changing the start/stop points
to build up the desired coverage versus optical phase delay.
Figure \ref{mtm_stroke} shows the concept.
A set of 6 different strokes, 
each of the same length but starting at different positions,
approximates the desired apodization
to achieve 95\% of the ideal observing efficiency.

%-------------------------------------------------------------------
% Appendix B: Comparison with FIRAS
%-------------------------------------------------------------------

\clearpage

\section{Comparison to FIRAS}
\label{sec:pixie_vs_firas}

The FIRAS spectrometer
was a cryogenic Martin-Puplett interferometer
whose two input ports compared signals from sky horn
to emission from an internal blackbody reference.
The sky horn could be blocked by a full-aperture blackbody
to provide an absolute reference.
Each of the two output ports
used a dichroic splitter at 600~GHz
to route the phase-delayed signal
to a high-frequency or low-frequency detector
\cite{firas_design_spie_1993}.
PIXIE improves the sensitivity of the COBE/FIRAS measurements
by a factor of more than 1000.  
We briefly review the principal design features 
leading to this improvement.

\begin{itemize}

\item{\bf Detector temperature:}
A liquid helium dewar cooled the FIRAS detectors to 1.5~K.
At this temperature, the dominant noise source
was phonon noise from the detectors,
producing a system NEP of 
$4 \times 10^{-15}~\rm{W}/\sqrt{{\rm Hz}}$.
An adiabatic demagnetization refrigerator 
maintains the PIXIE detectors at 0.1~K
to allow photon-limited NEP
$2.7 \times 10^{-16}\,{\rm W} /\sqrt{\rm Hz}$
($\S$\ref{sec:sensitivity}).

\item{\bf Etendue:}
Although each of the 4 FIRAS detectors had etendue
$A \Omega = 1.5$~cm$^2$~sr$^{-1}$,
poor noise performance in two detectors
meant that most of the sensitivity to CMB spectral distortions
came from a single detector
in the low-frequency dichroic channel at one of the two output ports.
The larger PIXIE detectors each provide etendue
$A \Omega = 4$~cm$^2$~sr$^{-1}$.
PIXIE will fly four such detectors,
each sensitive to the full optical passband.

\item{\bf Integration Time:}
The FIRAS mission ended after 10 months
when the liquid helium coolant ran out.
Only 1 month of this time was spent 
observing the external blackbody calibrator,
so that the sensitivity to sky spectra
was limited by the signal-to-noise ratio
of the calibration data.
Since the single sky horn could not directly
compare the sky to the external calibrator,
the resulting comparison of separate sky/calibrator data sets
incurred an additional $\sqrt{2}$ noise penalty.
PIXIE has no expendable cryogens
and can operate for extended periods.
The baseline mission devotes 10.8 months
to measurements of spectral distortions over the full sky,
during which
PIXIE's 2-beam design directly acquires
the sky/calibrator difference spectrum.

\item{\bf Interferogram Acquisition:}
FIRAS employed multiple scan modes to acquire the interferograms:
a ``short'' stroke with 512 samples over a 1.2~cm path difference 
vs a ``long'' stroke with 2048 samples over a 5.9~cm path difference.
Each stroke length was observed at 2 speeds:
a ``slow'' stroke of 0.783 cm/s vs a ``fast'' stroke of 1.175 cm/s.
The sensitivity for continuum sources such as the CMB
was dominated by the short/slow combination,
with little contribution from integration time
devoted to other modes.
PIXIE will divide its integration time
between two different stroke lengths,
with the bulk (70\%) devoted to shorter strokes
directed at the CMB continuum
and the remainder used to  obtain additional channels
at frequencies below 60~GHz
($\S$\ref{sec:sensitivity}).

\end{itemize}

\noindent
Table \ref{pixie_vs_firas_table} 
summarizes the relative improvements in sensitivity.

% -------------- Table C: PIXIE vs FIRAS  --------------
\begin{table}[t]
{
\caption{PIXIE vs FIRAS Sensitivity Improvements
\label{pixie_vs_firas_table}}
\begin{center}
\begin{tabular}{l c}
\hline 
Design Feature & Sensitivity Improvement	\\
\hline 
Detector Temperature		&	$\times$17	\\
Etendue				&	$\times$3	\\	
Integration Time		&	$\times$5	\\	
Interferogram Acquisition	&	$\times$4	\\	
\hline 
Total				&	$\times$1020	\\
\hline 
\end{tabular}
\end{center}
}
\end{table}
%------------------------------------------------------------

Although FIRAS was not limited by systematic errors,
the improved PIXIE sensitivity requires a corresponding
reduction in potential sources of systematic error.
Contributions to the PIXIE systematic error budget
are discussed in
\cite{
nagler_syserr_2015,
pixie_calibration,
pixie_syserr:2023}.
We briefly review the principal design elements
leading to a lower systematic error budget.

\begin{itemize}

\item{\bf Isothermal Operation:}
The bulk of the FIRAS instrument was cooled to 1.4~K 
by a superfluid liquid helium dewar,
with the internal and external calibrators maintained near 2.725~K.
The resulting 1.3~K gradient is a potential source for systematic error.
PIXIE maintains all elements except the detectors within 5~mK 
of the CMB temperature,
reducing the gradient source term by a factor of 265.

\item{\bf Blackbody Calibrator:}
FIRAS compared the sky to an external 
trumpet-mute calibrator
with emissivity $\epsilon > 0.99995$
(power reflection $R < -43$~dB)
\cite{firas_xcal_1999}.
The PIXIE calibrator uses an array of absorbing cones
with calculated reflectance
$R < -65$~dB 
\cite{pixie_calibration}.

\item{\bf Internal Reference:}
FIRAS's two input ports compared the signal from the sky horn
to an internal reference source
with emissivity  $\epsilon = 0.96$.
FIRAS data required corrections both for
reflections from the internal reference source
as well as temperature gradients
from heating at its tip
\cite{firas_calibration_1994}.
PIXIE has no such source,
and directly compares the sky to an external blackbody calibrator.

\item{\bf Cosmic Rays:}
Cosmic ray hits to the FIRAS detectors
were a dominant source of noise.
FIRAS co-added multiple interferogram scans
before downloading the data,
preventing cleaning of cosmic ray hits in the time-ordered data.
Despite the larger size of the PIXIE detectors,
the PIXIE open construction
has only 40\% of the FIRAS effective area for particle hits.
PIXIE downloads the full time-ordered data
for individual interferograms,
without onboard coaddition,
allowing cosmic ray cleaning.
After cleaning,
cosmic rays produce an approximate white-noise spectrum
a factor of 10 below the PIXIE photon noise
\cite{pixie_syserr:2023}.

\item{\bf Interferogram Acquisition:}
FIRAS acquired double-sided interferograms
so that the sky signal was  entirely in the real part of the Fourier transform,
but took data in only one mirror stroke direction.
PIXIE will similarly acquire double-sided interferograms,
but will take data in both directions
to obtain additional time-reversal symmetries.

\item{\bf Instrument Symmetry:}
The restricted volume within COBE's liquid helium dewar
required FIRAS to use a folded optical path
and a single sky horn.
PIXIE is fully symmetric,
with mirror-image beam-forming optics
illuminating a fully-symmetric
Martin-Puplett interferometer.
The multiple symmetries enable jackknife tests
to detect, model, and remove potential systematic errors
\cite{pixie_syserr:2023}.

\end{itemize}

%-------------------------------------------------------------------
% Appendix C: Mission noise curves
%-------------------------------------------------------------------

% For now, start appendices on new page
\clearpage

\section{Mission Noise Curves}
\label{sec:noise_curves}

The tables below contain the center frequencies,
sensitivity, and median noise level for a map pixel
for the baseline 2-year PIXIE mission.
The baseline 2-year mission 
assumes that 45\% of observing time is spent with the
calibrator deployed 
(sensitive to spectral distortions and polarization)
and
45\% with the calibrator stowed
(sensitive to polarization only).
Entries for polarization noise (Table \ref{pol_table})
include the polarization contribution from both modes.
Noise levels assume 30\% of the observing time in each mode
is spent with narrow synthesized channel width
(MTM long-stroke mode)
and 70\% with broad channel width
(MTM short-stroke mode).
See $\S$\ref{sec:sensitivity} for details.
The channel widths depend on the MTM mirror stroke
and can be changed (within hardware limits) throughout the mission.

Data in these tables are also available on the
Legacy Archive for Microwave Background Data (LAMBDA),
{\tt https://lambda.gsfc.nasa.gov/product/pixie/pixie\_baseline\_noise\_get.html}.

\clearpage

% -------------- Table C1: Spectral Distortion Sensitivity --------------
\begin{longtable}{|c c c || c c c|}
\caption{Mission Sensitivity to Spectral Distortions\label{sd_table}} \\
\hline
Center Frequency  &   Sensitivity   &   Map Noise	&
Center Frequency  &   Sensitivity   &   Map Noise	\\
   (GHz)	  &   (Jy sr$^{-1}~\sqrt{{\rm sec}}$	& Jy sr$^{-1}$ &
   (GHz)	  &   (Jy sr$^{-1}~\sqrt{{\rm sec}}$	& Jy sr$^{-1}$ \\
\hline 
\endfirsthead
\multicolumn{6}{c}%
{{\bfseries \tablename\ \thetable{}} -- continued from previous page} \\
\hline
Center Frequency  &   Sensitivity   &   Map Noise	&
Center Frequency  &   Sensitivity   &   Map Noise	\\
   (GHz)	  &   (Jy sr$^{-1}~\sqrt{{\rm sec}}$	& Jy sr$^{-1}$ &
   (GHz)	  &   (Jy sr$^{-1}~\sqrt{{\rm sec}}$	& Jy sr$^{-1}$ \\
\hline
\endhead
\multicolumn{6}{r}{{Continued on next page}} \\ 
\endfoot

\endlastfoot
  28.8     &   6.09E+04   &     2.53E+03     &      768.5     &   1.29E+04   &     5.37E+02     \\
  38.4     &   4.91E+04   &     2.04E+03     &      787.7     &   1.33E+04   &     5.53E+02     \\
  48.0     &   4.11E+04   &     1.71E+03     &      806.9     &   1.34E+04   &     5.59E+02     \\
  57.6     &   1.04E+04   &     4.33E+02     &      826.1     &   1.33E+04   &     5.55E+02     \\
  76.8     &   8.34E+03   &     3.47E+02     &      845.3     &   1.32E+04   &     5.47E+02     \\
  96.1     &   7.22E+03   &     3.00E+02     &      864.5     &   1.31E+04   &     5.46E+02     \\
 115.3     &   6.67E+03   &     2.78E+02     &      883.7     &   1.29E+04   &     5.39E+02     \\
 134.5     &   6.82E+03   &     2.84E+02     &      903.0     &   1.28E+04   &     5.34E+02     \\
 153.7     &   6.33E+03   &     2.63E+02     &      922.2     &   1.29E+04   &     5.35E+02     \\
 172.9     &   6.28E+03   &     2.61E+02     &      941.4     &   1.30E+04   &     5.41E+02     \\
 192.1     &   6.49E+03   &     2.70E+02     &      960.6     &   1.33E+04   &     5.51E+02     \\
 211.3     &   7.16E+03   &     2.98E+02     &      979.8     &   1.36E+04   &     5.66E+02     \\
 230.5     &   8.75E+03   &     3.64E+02     &      999.0     &   1.41E+04   &     5.86E+02     \\
 249.8     &   1.01E+04   &     4.19E+02     &     1018.2     &   1.46E+04   &     6.09E+02     \\
 269.0     &   1.14E+04   &     4.75E+02     &     1037.4     &   1.53E+04   &     6.36E+02     \\
 288.2     &   1.20E+04   &     4.98E+02     &     1056.7     &   1.59E+04   &     6.62E+02     \\
 307.4     &   1.17E+04   &     4.88E+02     &     1075.9     &   1.65E+04   &     6.85E+02     \\
 326.6     &   1.16E+04   &     4.82E+02     &     1095.1     &   1.69E+04   &     7.02E+02     \\
 345.8     &   1.03E+04   &     4.28E+02     &     1114.3     &   1.72E+04   &     7.14E+02     \\
 365.0     &   1.02E+04   &     4.25E+02     &     1133.5     &   1.73E+04   &     7.19E+02     \\
 384.2     &   9.56E+03   &     3.98E+02     &     1152.7     &   1.73E+04   &     7.18E+02     \\
 403.4     &   9.16E+03   &     3.81E+02     &     1171.9     &   1.72E+04   &     7.14E+02     \\
 422.7     &   8.84E+03   &     3.68E+02     &     1191.1     &   1.71E+04   &     7.11E+02     \\
 441.9     &   8.74E+03   &     3.64E+02     &     1210.3     &   1.71E+04   &     7.12E+02     \\
 461.1     &   9.42E+03   &     3.92E+02     &     1229.6     &   1.73E+04   &     7.18E+02     \\
 480.3     &   9.68E+03   &     4.03E+02     &     1248.8     &   1.76E+04   &     7.31E+02     \\
 499.5     &   1.10E+04   &     4.56E+02     &     1268.0     &   1.81E+04   &     7.52E+02     \\
 518.7     &   1.12E+04   &     4.64E+02     &     1287.2     &   1.87E+04   &     7.79E+02     \\
 537.9     &   1.15E+04   &     4.79E+02     &     1306.4     &   1.95E+04   &     8.09E+02     \\
 557.1     &   1.15E+04   &     4.80E+02     &     1325.6     &   2.02E+04   &     8.41E+02     \\
 576.4     &   1.13E+04   &     4.72E+02     &     1344.8     &   2.08E+04   &     8.65E+02     \\
 595.6     &   1.10E+04   &     4.59E+02     &     1364.0     &   2.14E+04   &     8.92E+02     \\
 614.8     &   1.10E+04   &     4.56E+02     &     1383.3     &   2.19E+04   &     9.13E+02     \\
 634.0     &   1.07E+04   &     4.46E+02     &     1402.5     &   2.22E+04   &     9.25E+02     \\
 653.2     &   1.05E+04   &     4.37E+02     &     1421.7     &   2.24E+04   &     9.31E+02     \\
 672.4     &   1.04E+04   &     4.31E+02     &     1440.9     &   2.24E+04   &     9.32E+02     \\
 691.6     &   1.04E+04   &     4.33E+02     &     1460.1     &   2.24E+04   &     9.33E+02     \\
 710.8     &   1.08E+04   &     4.48E+02     &     1479.3     &   2.25E+04   &     9.36E+02     \\
 730.1     &   1.15E+04   &     4.77E+02     &     1498.5     &   2.27E+04   &     9.46E+02     \\
 749.3     &   1.23E+04   &     5.10E+02     &     1517.7     &   2.31E+04   &     9.62E+02     \\
\hline
\newpage
1536.9     &   2.37E+04   &     9.87E+02     &     2305.4     &   5.50E+04   &     2.29E+03     \\
1556.2     &   2.45E+04   &     1.02E+03     &     2324.6     &   5.60E+04   &     2.33E+03     \\
1575.4     &   2.54E+04   &     1.05E+03     &     2343.8     &   5.70E+04   &     2.37E+03     \\
1594.6     &   2.61E+04   &     1.09E+03     &     2363.1     &   5.83E+04   &     2.42E+03     \\
1613.8     &   2.71E+04   &     1.13E+03     &     2382.3     &   5.99E+04   &     2.49E+03     \\
1633.0     &   2.79E+04   &     1.16E+03     &     2401.5     &   6.18E+04   &     2.57E+03     \\
1652.2     &   2.87E+04   &     1.19E+03     &     2420.7     &   6.38E+04   &     2.66E+03     \\
1671.4     &   2.92E+04   &     1.21E+03     &     2439.9     &   6.59E+04   &     2.74E+03     \\
1690.6     &   2.95E+04   &     1.23E+03     &     2459.1     &   6.78E+04   &     2.82E+03     \\
1709.9     &   2.97E+04   &     1.23E+03     &     2478.3     &   6.95E+04   &     2.89E+03     \\
1729.1     &   2.98E+04   &     1.24E+03     &     2497.5     &   7.09E+04   &     2.95E+03     \\
1748.3     &   3.00E+04   &     1.25E+03     &     2516.8     &   7.19E+04   &     2.99E+03     \\
1767.5     &   3.03E+04   &     1.26E+03     &     2536.0     &   7.28E+04   &     3.03E+03     \\
1786.7     &   3.08E+04   &     1.28E+03     &     2555.2     &   7.36E+04   &     3.06E+03     \\
1805.9     &   3.15E+04   &     1.31E+03     &     2574.4     &   7.48E+04   &     3.11E+03     \\
1825.1     &   3.24E+04   &     1.35E+03     &     2593.6     &   7.58E+04   &     3.15E+03     \\
1844.3     &   3.33E+04   &     1.39E+03     &     2612.8     &   7.71E+04   &     3.21E+03     \\
1863.6     &   3.45E+04   &     1.44E+03     &     2632.0     &   7.88E+04   &     3.28E+03     \\
1882.8     &   3.57E+04   &     1.49E+03     &     2651.2     &   8.08E+04   &     3.36E+03     \\
1902.0     &   3.69E+04   &     1.54E+03     &     2670.4     &   8.32E+04   &     3.46E+03     \\
1921.2     &   3.79E+04   &     1.58E+03     &     2689.7     &   8.59E+04   &     3.57E+03     \\
1940.4     &   3.87E+04   &     1.61E+03     &     2708.9     &   8.86E+04   &     3.69E+03     \\
1959.6     &   3.93E+04   &     1.63E+03     &     2728.1     &   9.12E+04   &     3.79E+03     \\
1978.8     &   3.97E+04   &     1.65E+03     &     2747.3     &   9.36E+04   &     3.89E+03     \\
1998.0     &   4.00E+04   &     1.66E+03     &     2766.5     &   9.56E+04   &     3.98E+03     \\
2017.2     &   4.03E+04   &     1.68E+03     &     2785.7     &   9.72E+04   &     4.04E+03     \\
2036.5     &   4.07E+04   &     1.69E+03     &     2804.9     &   9.90E+04   &     4.12E+03     \\
2055.7     &   4.14E+04   &     1.72E+03     &     2824.1     &   1.00E+05   &     4.17E+03     \\
2074.9     &   4.22E+04   &     1.76E+03     &     2843.4     &   1.01E+05   &     4.22E+03     \\
2094.1     &   4.33E+04   &     1.80E+03     &     2862.6     &   1.03E+05   &     4.28E+03     \\
2113.3     &   4.46E+04   &     1.85E+03     &     2881.8     &   1.04E+05   &     4.35E+03     \\
2132.5     &   4.61E+04   &     1.92E+03     &     2901.0     &   1.07E+05   &     4.43E+03     \\
2151.7     &   4.76E+04   &     1.98E+03     &     2920.2     &   1.09E+05   &     4.54E+03     \\
2170.9     &   4.92E+04   &     2.05E+03     &     2939.4     &   1.12E+05   &     4.67E+03     \\
2190.2     &   5.06E+04   &     2.11E+03     &     2958.6     &   1.16E+05   &     4.81E+03     \\
2209.4     &   5.18E+04   &     2.15E+03     &     2977.8     &   1.19E+05   &     4.96E+03     \\
2228.6     &   5.27E+04   &     2.19E+03     &     2997.1     &   1.23E+05   &     5.11E+03     \\
2247.8     &   5.33E+04   &     2.22E+03     &     3016.3     &   1.26E+05   &     5.25E+03     \\
2267.0     &   5.39E+04   &     2.24E+03     &     3035.5     &   1.29E+05   &     5.37E+03     \\
2286.2     &   5.44E+04   &     2.26E+03     &     3054.7     &   1.32E+05   &     5.48E+03     \\
\hline
\newpage
3073.9     &   1.34E+05   &     5.57E+03     &     3842.4     &   3.17E+05   &     1.32E+04     \\
3093.1     &   1.36E+05   &     5.65E+03     &     3861.6     &   3.24E+05   &     1.35E+04     \\
3112.3     &   1.38E+05   &     5.72E+03     &     3880.8     &   3.31E+05   &     1.38E+04     \\
3131.5     &   1.39E+05   &     5.80E+03     &     3900.0     &   3.37E+05   &     1.40E+04     \\
3150.7     &   1.42E+05   &     5.89E+03     &     3919.2     &   3.42E+05   &     1.42E+04     \\
3170.0     &   1.44E+05   &     6.01E+03     &     3938.4     &   3.47E+05   &     1.44E+04     \\
3189.2     &   1.48E+05   &     6.15E+03     &     3957.6     &   3.52E+05   &     1.47E+04     \\
3208.4     &   1.52E+05   &     6.31E+03     &     3976.9     &   3.59E+05   &     1.49E+04     \\
3227.6     &   1.56E+05   &     6.50E+03     &     3996.1     &   3.66E+05   &     1.52E+04     \\
3246.8     &   1.61E+05   &     6.69E+03     &     4015.3     &   3.74E+05   &     1.56E+04     \\
3266.0     &   1.66E+05   &     6.89E+03     &     4034.5     &   3.83E+05   &     1.59E+04     \\
3285.2     &   1.70E+05   &     7.08E+03     &     4053.7     &   3.93E+05   &     1.64E+04     \\
3304.4     &   1.74E+05   &     7.24E+03     &     4072.9     &   4.04E+05   &     1.68E+04     \\
3323.7     &   1.78E+05   &     7.40E+03     &     4092.1     &   4.16E+05   &     1.73E+04     \\
3342.9     &   1.81E+05   &     7.54E+03     &     4111.3     &   4.27E+05   &     1.78E+04     \\
3362.1     &   1.84E+05   &     7.65E+03     &     4130.6     &   4.37E+05   &     1.82E+04     \\
3381.3     &   1.86E+05   &     7.76E+03     &     4149.8     &   4.47E+05   &     1.86E+04     \\
3400.5     &   1.89E+05   &     7.87E+03     &     4169.0     &   4.55E+05   &     1.89E+04     \\
3419.7     &   1.92E+05   &     7.99E+03     &     4188.2     &   4.62E+05   &     1.92E+04     \\
3438.9     &   1.96E+05   &     8.14E+03     &     4207.4     &   4.70E+05   &     1.95E+04     \\
3458.1     &   2.00E+05   &     8.32E+03     &     4226.6     &   4.77E+05   &     1.98E+04     \\
3477.3     &   2.05E+05   &     8.53E+03     &     4245.8     &   4.85E+05   &     2.02E+04     \\
3496.6     &   2.11E+05   &     8.77E+03     &     4265.0     &   4.94E+05   &     2.06E+04     \\
3515.8     &   2.17E+05   &     9.03E+03     &     4284.2     &   5.05E+05   &     2.10E+04     \\
3535.0     &   2.23E+05   &     9.26E+03     &     4303.5     &   5.16E+05   &     2.15E+04     \\
3554.2     &   2.29E+05   &     9.53E+03     &     4322.7     &   5.30E+05   &     2.20E+04     \\
3573.4     &   2.35E+05   &     9.77E+03     &     4341.9     &   5.44E+05   &     2.26E+04     \\
3592.6     &   2.40E+05   &     1.00E+04     &     4361.1     &   5.60E+05   &     2.33E+04     \\
3611.8     &   2.45E+05   &     1.02E+04     &     4380.3     &   5.75E+05   &     2.39E+04     \\
3631.0     &   2.49E+05   &     1.04E+04     &     4399.5     &   5.89E+05   &     2.45E+04     \\
3650.3     &   2.53E+05   &     1.05E+04     &     4418.7     &   6.02E+05   &     2.51E+04     \\
3669.5     &   2.56E+05   &     1.07E+04     &     4437.9     &   6.14E+05   &     2.55E+04     \\
3688.7     &   2.60E+05   &     1.08E+04     &     4457.2     &   6.24E+05   &     2.60E+04     \\
3707.9     &   2.65E+05   &     1.10E+04     &     4476.4     &   6.34E+05   &     2.64E+04     \\
3727.1     &   2.70E+05   &     1.13E+04     &     4495.6     &   6.44E+05   &     2.68E+04     \\
3746.3     &   2.77E+05   &     1.15E+04     &     4514.8     &   6.56E+05   &     2.73E+04     \\
3765.5     &   2.84E+05   &     1.18E+04     &     4534.0     &   6.67E+05   &     2.78E+04     \\
3784.7     &   2.92E+05   &     1.21E+04     &     4553.2     &   6.80E+05   &     2.83E+04     \\
3803.9     &   3.00E+05   &     1.25E+04     &     4572.4     &   6.96E+05   &     2.89E+04     \\
3823.2     &   3.09E+05   &     1.28E+04     &     4591.6     &   7.13E+05   &     2.97E+04     \\
\hline
\newpage
4610.8     &   7.32E+05   &     3.05E+04     &     5379.3     &   1.68E+06   &     7.00E+04     \\
4630.1     &   7.52E+05   &     3.13E+04     &     5398.5     &   1.72E+06   &     7.15E+04     \\
4649.3     &   7.72E+05   &     3.21E+04     &     5417.7     &   1.76E+06   &     7.32E+04     \\
4668.5     &   7.92E+05   &     3.29E+04     &     5437.0     &   1.81E+06   &     7.51E+04     \\
4687.7     &   8.10E+05   &     3.37E+04     &     5456.2     &   1.85E+06   &     7.70E+04     \\
4706.9     &   8.26E+05   &     3.44E+04     &     5475.4     &   1.90E+06   &     7.90E+04     \\
4726.1     &   8.41E+05   &     3.50E+04     &     5494.6     &   1.94E+06   &     8.08E+04     \\
4745.3     &   8.55E+05   &     3.56E+04     &     5513.8     &   1.99E+06   &     8.26E+04     \\
4764.5     &   8.70E+05   &     3.62E+04     &     5533.0     &   2.03E+06   &     8.43E+04     \\
4783.8     &   8.84E+05   &     3.68E+04     &     5552.2     &   2.07E+06   &     8.59E+04     \\
4803.0     &   8.99E+05   &     3.74E+04     &     5571.4     &   2.10E+06   &     8.73E+04     \\
4822.2     &   9.16E+05   &     3.81E+04     &     5590.7     &   2.13E+06   &     8.87E+04     \\
4841.4     &   9.36E+05   &     3.89E+04     &     5609.9     &   2.17E+06   &     9.01E+04     \\
4860.6     &   9.58E+05   &     3.99E+04     &     5629.1     &   2.20E+06   &     9.17E+04     \\
4879.8     &   9.83E+05   &     4.09E+04     &     5648.3     &   2.25E+06   &     9.34E+04     \\
4899.0     &   1.01E+06   &     4.20E+04     &     5667.5     &   2.29E+06   &     9.54E+04     \\
4918.2     &   1.04E+06   &     4.31E+04     &     5686.7     &   2.35E+06   &     9.76E+04     \\
4937.4     &   1.06E+06   &     4.42E+04     &     5705.9     &   2.40E+06   &     1.00E+05     \\
4956.7     &   1.09E+06   &     4.52E+04     &     5725.1     &   2.46E+06   &     1.03E+05     \\
4975.9     &   1.11E+06   &     4.62E+04     &     5744.3     &   2.52E+06   &     1.05E+05     \\
4995.1     &   1.13E+06   &     4.70E+04     &     5763.6     &   2.58E+06   &     1.07E+05     \\
5014.3     &   1.15E+06   &     4.79E+04     &     5782.8     &   2.64E+06   &     1.10E+05     \\
5033.5     &   1.17E+06   &     4.87E+04     &     5802.0     &   2.70E+06   &     1.12E+05     \\
5052.7     &   1.19E+06   &     4.95E+04     &     5821.2     &   2.75E+06   &     1.14E+05     \\
5071.9     &   1.21E+06   &     5.03E+04     &     5840.4     &   2.80E+06   &     1.16E+05     \\
5091.1     &   1.23E+06   &     5.12E+04     &     5859.6     &   2.84E+06   &     1.18E+05     \\
5110.4     &   1.26E+06   &     5.22E+04     &     5878.8     &   2.89E+06   &     1.20E+05     \\
5129.6     &   1.28E+06   &     5.34E+04     &     5898.0     &   2.93E+06   &     1.22E+05     \\
5148.8     &   1.32E+06   &     5.48E+04     &     5917.3     &   2.99E+06   &     1.24E+05     \\
5168.0     &   1.35E+06   &     5.62E+04     &     5936.5     &   3.05E+06   &     1.27E+05     \\
5187.2     &   1.39E+06   &     5.77E+04     &     5955.7     &   3.11E+06   &     1.30E+05     \\
5206.4     &   1.42E+06   &     5.92E+04     &     5974.9     &   3.18E+06   &     1.32E+05     \\
5225.6     &   1.46E+06   &     6.06E+04     &     5994.1     &   3.41E+06   &     1.42E+05     \\
5244.8     &   1.49E+06   &6.19E+04	& \multicolumn{3}{c|}{ } \\
5264.1     &   1.52E+06   &6.31E+04	& \multicolumn{3}{c|}{ } \\
5283.3     &   1.54E+06   &6.43E+04	& \multicolumn{3}{c|}{ } \\
5302.5     &   1.57E+06   &6.53E+04	& \multicolumn{3}{c|}{ } \\
5321.7     &   1.59E+06   &6.63E+04	& \multicolumn{3}{c|}{ } \\
5340.9     &   1.62E+06   &6.74E+04	& \multicolumn{3}{c|}{ } \\
5360.1     &   1.65E+06   &6.86E+04	& \multicolumn{3}{c|}{ } \\
\hline
\end{longtable}
%------------------------------------------------------------

% -------------- Table C2: Polarization Sensitivity --------------

\begin{longtable}{|c c c || c c c|}
\caption{Mission Sensitivity to Polarizations\label{pol_table}} \\
\hline
Center Frequency  &   Sensitivity   &   Map Noise	&
Center Frequency  &   Sensitivity   &   Map Noise	\\
   (GHz)	  &   (Jy sr$^{-1}~\sqrt{{\rm sec}}$	& Jy sr$^{-1}$ &
   (GHz)	  &   (Jy sr$^{-1}~\sqrt{{\rm sec}}$	& Jy sr$^{-1}$ \\
\hline
\endfirsthead
\multicolumn{6}{c}%
{{\bfseries \tablename\ \thetable{}} -- continued from previous page} \\
\hline
Center Frequency  &   Sensitivity   &   Map Noise	&
Center Frequency  &   Sensitivity   &   Map Noise	\\
   (GHz)	  &   (Jy sr$^{-1}~\sqrt{{\rm sec}}$	& Jy sr$^{-1}$ &
   (GHz)	  &   (Jy sr$^{-1}~\sqrt{{\rm sec}}$	& Jy sr$^{-1}$ \\
\hline
\endhead
\multicolumn{6}{r}{{Continued on next page}} \\ 
\endfoot

\endlastfoot
  28.8     &   8.61E+04   &     3.58E+03     &     1421.7     &   8.30E+03   &     3.45E+02     \\
  38.4     &   6.95E+04   &     2.89E+03     &     1460.1     &   8.32E+03   &     3.46E+02     \\
  48.0     &   5.82E+04   &     2.42E+03     &     1498.5     &   8.43E+03   &     3.51E+02     \\
  57.6     &   9.62E+03   &     4.00E+02     &     1536.9     &   8.79E+03   &     3.66E+02     \\
  76.8     &   7.93E+03   &     3.30E+02     &     1575.4     &   9.39E+03   &     3.91E+02     \\
  96.1     &   6.85E+03   &     2.85E+02     &     1613.8     &   1.00E+04   &     4.17E+02     \\
 115.3     &   2.57E+03   &     1.07E+02     &     1652.2     &   1.06E+04   &     4.42E+02     \\
 153.7     &   2.36E+03   &     9.81E+01     &     1690.6     &   1.09E+04   &     4.55E+02     \\
 192.1     &   2.41E+03   &     1.00E+02     &     1729.1     &   1.11E+04   &     4.60E+02     \\
 230.5     &   3.10E+03   &     1.29E+02     &     1767.5     &   1.12E+04   &     4.68E+02     \\
 269.0     &   4.20E+03   &     1.75E+02     &     1805.9     &   1.17E+04   &     4.86E+02     \\
 307.4     &   4.36E+03   &     1.81E+02     &     1844.3     &   1.24E+04   &     5.14E+02     \\
 345.8     &   3.85E+03   &     1.60E+02     &     1882.8     &   1.32E+04   &     5.51E+02     \\
 384.2     &   3.56E+03   &     1.48E+02     &     1921.2     &   1.41E+04   &     5.85E+02     \\
 422.7     &   3.29E+03   &     1.37E+02     &     1959.6     &   1.46E+04   &     6.06E+02     \\
 461.1     &   3.36E+03   &     1.40E+02     &     1998.0     &   1.48E+04   &     6.17E+02     \\
 499.5     &   3.89E+03   &     1.62E+02     &     2036.5     &   1.51E+04   &     6.29E+02     \\
 537.9     &   4.26E+03   &     1.77E+02     &     2074.9     &   1.57E+04   &     6.52E+02     \\
 576.4     &   4.21E+03   &     1.75E+02     &     2113.3     &   1.65E+04   &     6.87E+02     \\
 614.8     &   4.07E+03   &     1.69E+02     &     2151.7     &   1.76E+04   &     7.34E+02     \\
 653.2     &   3.90E+03   &     1.62E+02     &     2190.2     &   1.87E+04   &     7.80E+02     \\
 691.6     &   3.86E+03   &     1.61E+02     &     2228.6     &   1.95E+04   &     8.12E+02     \\
 730.1     &   4.24E+03   &     1.76E+02     &     2267.0     &   2.00E+04   &     8.31E+02     \\
 768.5     &   4.78E+03   &     1.99E+02     &     2305.4     &   2.04E+04   &     8.49E+02     \\
 806.9     &   4.98E+03   &     2.07E+02     &     2343.8     &   2.11E+04   &     8.79E+02     \\
 845.3     &   4.88E+03   &     2.03E+02     &     2382.3     &   2.22E+04   &     9.24E+02     \\
 883.7     &   4.81E+03   &     2.00E+02     &     2420.7     &   2.37E+04   &     9.84E+02     \\
 922.2     &   4.78E+03   &     1.99E+02     &     2459.1     &   2.51E+04   &     1.05E+03     \\
 960.6     &   4.91E+03   &     2.04E+02     &     2497.5     &   2.63E+04   &     1.09E+03     \\
 999.0     &   5.22E+03   &     2.17E+02     &     2536.0     &   2.70E+04   &     1.12E+03     \\
1037.4     &   5.66E+03   &     2.35E+02     &     2574.4     &   2.77E+04   &     1.15E+03     \\
1075.9     &   6.10E+03   &     2.54E+02     &     2612.8     &   2.86E+04   &     1.19E+03     \\
1114.3     &   6.36E+03   &     2.64E+02     &     2651.2     &   3.00E+04   &     1.25E+03     \\
1152.7     &   6.40E+03   &     2.66E+02     &     2689.7     &   3.18E+04   &     1.32E+03     \\
1191.1     &   6.34E+03   &     2.64E+02     &     2728.1     &   3.38E+04   &     1.41E+03     \\
1229.6     &   6.41E+03   &     2.66E+02     &     2766.5     &   3.54E+04   &     1.47E+03     \\
1268.0     &   6.70E+03   &     2.79E+02     &     2804.9     &   3.67E+04   &     1.53E+03     \\
1306.4     &   7.21E+03   &     3.00E+02     &     2843.4     &   3.76E+04   &     1.56E+03     \\
1344.8     &   7.71E+03   &     3.21E+02     &     2881.8     &   3.87E+04   &     1.61E+03     \\
1383.3     &   8.13E+03   &     3.38E+02     &     2920.2     &   4.05E+04   &     1.68E+03     \\
\hline
\newpage
2958.6     &   4.29E+04   &     1.78E+03     &     4495.6     &   2.39E+05   &     9.93E+03     \\
2997.1     &   4.55E+04   &     1.89E+03     &     4534.0     &   2.47E+05   &     1.03E+04     \\
3035.5     &   4.78E+04   &     1.99E+03     &     4572.4     &   2.58E+05   &     1.07E+04     \\
3073.9     &   4.96E+04   &     2.07E+03     &     4610.8     &   2.71E+05   &     1.13E+04     \\
3112.3     &   5.10E+04   &     2.12E+03     &     4649.3     &   2.86E+05   &     1.19E+04     \\
3150.7     &   5.25E+04   &     2.18E+03     &     4687.7     &   3.00E+05   &     1.25E+04     \\
3189.2     &   5.48E+04   &     2.28E+03     &     4726.1     &   3.12E+05   &     1.30E+04     \\
3227.6     &   5.79E+04   &     2.41E+03     &     4764.5     &   3.23E+05   &     1.34E+04     \\
3266.0     &   6.14E+04   &     2.55E+03     &     4803.0     &   3.33E+05   &     1.39E+04     \\
3304.4     &   6.45E+04   &     2.69E+03     &     4841.4     &   3.47E+05   &     1.44E+04     \\
3342.9     &   6.72E+04   &     2.79E+03     &     4879.8     &   3.64E+05   &     1.52E+04     \\
3381.3     &   6.92E+04   &     2.88E+03     &     4918.2     &   3.84E+05   &     1.60E+04     \\
3419.7     &   7.12E+04   &     2.96E+03     &     4956.7     &   4.03E+05   &     1.68E+04     \\
3458.1     &   7.41E+04   &     3.08E+03     &     4995.1     &   4.19E+05   &     1.74E+04     \\
3496.6     &   7.81E+04   &     3.25E+03     &     5033.5     &   4.34E+05   &     1.81E+04     \\
3535.0     &   8.25E+04   &     3.43E+03     &     5071.9     &   4.48E+05   &     1.86E+04     \\
3573.4     &   8.71E+04   &     3.62E+03     &     5110.4     &   4.66E+05   &     1.94E+04     \\
3611.8     &   9.08E+04   &     3.78E+03     &     5148.8     &   4.88E+05   &     2.03E+04     \\
3650.3     &   9.37E+04   &     3.90E+03     &     5187.2     &   5.14E+05   &     2.14E+04     \\
3688.7     &   9.65E+04   &     4.01E+03     &     5225.6     &   5.40E+05   &     2.25E+04     \\
3727.1     &   1.00E+05   &     4.17E+03     &     5264.1     &   5.63E+05   &     2.34E+04     \\
3765.5     &   1.05E+05   &     4.39E+03     &     5302.5     &   5.82E+05   &     2.42E+04     \\
3803.9     &   1.11E+05   &     4.63E+03     &     5340.9     &   6.01E+05   &     2.50E+04     \\
3842.4     &   1.17E+05   &     4.89E+03     &     5379.3     &   6.24E+05   &     2.59E+04     \\
3880.8     &   1.23E+05   &     5.11E+03     &     5417.7     &   6.52E+05   &     2.71E+04     \\
3919.2     &   1.27E+05   &     5.28E+03     &     5456.2     &   6.86E+05   &     2.86E+04     \\
3957.6     &   1.31E+05   &     5.44E+03     &     5494.6     &   7.20E+05   &     2.99E+04     \\
3996.1     &   1.36E+05   &     5.64E+03     &     5533.0     &   7.52E+05   &     3.13E+04     \\
4034.5     &   1.42E+05   &     5.91E+03     &     5571.4     &   7.78E+05   &     3.24E+04     \\
4072.9     &   1.50E+05   &     6.23E+03     &     5609.9     &   8.04E+05   &     3.34E+04     \\
4111.3     &   1.58E+05   &     6.58E+03     &     5648.3     &   8.33E+05   &     3.46E+04     \\
4149.8     &   1.66E+05   &     6.89E+03     &     5686.7     &   8.70E+05   &     3.62E+04     \\
4188.2     &   1.71E+05   &     7.13E+03     &     5725.1     &   9.13E+05   &     3.80E+04     \\
4226.6     &   1.77E+05   &     7.35E+03     &     5763.6     &   9.58E+05   &     3.98E+04     \\
4265.0     &   1.83E+05   &     7.62E+03     &     5802.0     &   1.00E+06   &     4.16E+04     \\
4303.5     &   1.91E+05   &     7.96E+03     &     5840.4     &   1.04E+06   &     4.31E+04     \\
4341.9     &   2.02E+05   &     8.39E+03     &     5878.8     &   1.07E+06   &     4.45E+04     \\
4380.3     &   2.13E+05   &     8.86E+03     &     5917.3     &   1.11E+06   &     4.61E+04     \\
4418.7     &   2.23E+05   &     9.29E+03     &     5955.7     &   1.15E+06   &     4.80E+04     \\
4457.2     &   2.31E+05   &     9.63E+03     &     5994.1     &   1.25E+06   &     5.22E+04     \\
\hline
\end{longtable}

%------------------------------------------------------------

% -------------- References --------------

% For now, put references on a new page
\clearpage

\bibliographystyle{JHEP}		%>>>> Force bibtex to use JHEP.bst
\bibliography{pixie_2021_mission}

% That's all there is, kiddies.  Ride off into the sunset!
\end{spacing}
\end{document}